\documentclass[prd,aps,11pt,
nofootinbib,tightenlines,
superscriptaddress,preprintnumbers]{revtex4}

\usepackage{graphicx}
\usepackage{amsmath}
\usepackage{amssymb}

\newcommand{\beq}{\begin{equation}}
\newcommand{\eeq}{\end{equation}}
\newcommand{\bqa}{\begin{eqnarray}}
\newcommand{\eqa}{\end{eqnarray}}
\newcommand{\be}{\begin{equation}}
\newcommand{\ee}{\end{equation}}
\newcommand{\bea}{\begin{eqnarray}}
\newcommand{\eea}{\end{eqnarray}}
\newcommand{\nn}{\nonumber}
\newcommand{\0}{\over }

\newcommand{\2}{{1\over2}}

\newcommand{\6}{\partial }
\newcommand{\9}{}

\newcommand{\tr}{\,{\rm tr}\,}

\newcommand{\cH}{\cosh(y-\eta) }
\newcommand{\sH}{\sinh(y-\eta) }
\newcommand{\tH}{\tanh(y-\eta) }
\newcommand{\atanh}{{\rm atanh}\,}

\newcommand{\bt}{\bar\tau }

\begin{document}

\title{Instabilities of an anisotropically expanding
non-Abelian plasma:\\ 1D+3V discretized hard-loop simulations}

\preprint{TUW-08-05, NSF-KITP-08-01}

\author{Anton Rebhan}
\affiliation{Institut f\"ur Theoretische Physik, Technische Universit\"at Wien,
        Wiedner Hauptstrasse 8-10, A-1040 Vienna, Austria}
\author{Michael Strickland}
\affiliation{Frankfurt Institute for Advanced Studies,
Johann Wolfgang Goethe University,\\
Max-von-Laue-Str.\ 1,
D-60438 Frankfurt am Main, Germany}
\affiliation{Kavli Institute for Theoretical Physics,
University of California Santa Barbara,\\
Santa Barbara, CA 93106, USA}
\author{Maximilian Attems}
\affiliation{Institut f\"ur Theoretische Physik, Technische Universit\"at Wien,
        Wiedner Hauptstrasse 8-10, A-1040 Vienna, Austria}

\date{\today}


\begin{abstract}
Non-Abelian plasma instabilities play a crucial 
role in the nonequilibrium dynamics of a weakly coupled quark-gluon 
plasma and they importantly modify the standard perturbative bottom-up 
thermalization scenario in heavy-ion collisions. Using the 
auxiliary-field formulation of the hard-loop effective 
theory, we study numerically the real time evolution of instabilities 
in an anisotropic collisionless Yang-Mills plasma expanding 
longitudinally in free streaming. In this first real-time lattice 
simulation we consider the most unstable modes, long-wavelength 
coherent color fields that are constant in transverse directions and 
which therefore are effectively 1+1-dimensional in spacetime, except 
for the auxiliary fields which also depend on discretized momentum 
rapidity and transverse velocity components. We reproduce the semi-analytical 
results obtained previously for the Abelian regime and we 
determine the nonlinear effects which occur when the instabilities have 
grown such that non-Abelian interactions become important.
\end{abstract}

\maketitle

\section{Introduction}

The experimental results obtained at the Relativistic Heavy Ion Collider
RHIC \cite{Tannenbaum:2006ch}
and their good agreement with hydrodynamical simulations  with
extremely early thermalization and a low shear viscosity 
\cite{Teaney:2003kp,Romatschke:2007mq,Song:2007ux} close
to a conjectured lower quantum theoretical bound \cite{Kovtun:2004de}
are now widely interpreted as evidence that the hypothetical quark-gluon matter
produced at RHIC is strongly interacting and very far from a
perturbatively accessible regime. Indeed, perturbative approaches
like that of the original
bottom-up thermalization scenario \cite{Wong:1996va,Wong:1997dv,Baier:2000sb} do not seem to
be able to come close to explaining the fast apparent thermalization.
However, as pointed out first by Ref.~\cite{Arnold:2003rq}, the original
bottom-up scenario is qualitatively changed by the inevitable
presence of non-Abelian (chromo-Weibel)
plasma instabilities \cite{Mrowczynski:1988dz,Pokrovsky:1988bm,Mrowczynski:1993qm}
in a weakly coupled quark-gluon plasma
with momentum-space anisotropy,
although it is still an open theoretical question how the bottom-up scenario
will have to be modified, even
at asymptotically weak coupling and in the first stage of
the bottom-up scenario 
\cite{Mueller:2005un,Bodeker:2005nv,Arnold:2005qs,Mueller:2006up,Arnold:2007cg}.
Non-Abelian plasma instabilities have moreover been argued to
importantly modify weak-coupling results on the shear viscosity
to anomalously low values \cite{Asakawa:2006tc}.
Even if the quark-gluon matter produced at RHIC may be too close
to the deconfinement phase transition for any extrapolations of
weak-coupling results, it is clearly necessary to better understand
the latter and how they differ from other approaches. Finally,
it may be the case that the higher energies to be reached at upcoming 
heavy-ion collider experiments
at the Large Hadron Collider (LHC) open the window to
the specific collective phenomena of a weakly coupled quark-gluon plasma,
such as non-Abelian plasma instabilities.

In this paper we shall discuss only the theoretically clean situation
at asymptotically weak coupling and the dynamical evolution of
non-Abelian plasma instabilities in a collisionless plasma with
long-wavelength color fields. Any amount of momentum anisotropy
in the distribution of the (high-momentum) plasma particles leads
to chromomagnetic instabilities, which
in the weak-field situation are straightforward generalizations of
the Abelian Weibel instabilities \cite{Weibel:1959} 
and whose dispersion laws have been
worked out for specific cases of a stationary anisotropic plasma
in Ref.~\cite{Romatschke:2003ms,Romatschke:2004jh,Schenke:2006xu,Schenke:2006fz}.
In an Abelian plasma, the Weibel instabilities grow exponentially
until they are large enough to modify the distribution of the hard particles
and give rise to their fast isotropization. In a weakly coupled non-Abelian
plasma, the situation is more complicated because the long-wavelength
color fields have nonlinear self-interactions before they reach the
size where fast isotropization occurs. The first numerical simulations
\cite{Rebhan:2004ur}
of non-Abelian plasma instabilities using the systematic framework
of the hard-loop effective theory \cite{Blaizot:2001nr,Pisarski:1997cp,Mrowczynski:2000ed,Mrowczynski:2004kv} have
concentrated on the most unstable modes which are constant
in the directions transverse to the direction of momentum anisotropy.
It was found that such configurations experience a certain amount of
Abelianization over domains of finite size
when they enter the nonlinear regime, which allows them to
continue an exponential growth out of the hard-loop regime,
confirming essentially the conjecture of Ref.~\cite{Arnold:2004ih}
formed from numerical studies of a toy model which showed
virtually complete Abelianization.
In spacetime, the corresponding evolution equations are
1+1 dimensional, which in the hard-loop effective theory
are coupled to auxiliary fields that depend on the three-dimensional
velocity of the hard particles,
so that in conventional plasma physics these simulations would be termed 1D+3V.
Fully 3+1 dimensional simulations (3D+3V) later showed however that
more generic field configurations in a plasma with fixed (moderate)
momentum space anisotropy do not continue to grow exponentially
in the strong-field regime, but enter a linear-growth phase \cite{Arnold:2005vb,Rebhan:2005re}
by the formation of a cascade which pumps the growing energy in the infrared
modes into higher-momentum modes \cite{Arnold:2005ef,Arnold:2005qs}.
The recent simulations of Ref.~\cite{Bodeker:2007fw}
however found a continued exponential
growth of initially small perturbations
in the case of very strong momentum anisotropy. A very strong anisotropy
(if not the requirement of initially small fluctuations \cite{Arnold:2007cg}) is
of particular interest for heavy-ion collision 
where in a weak coupling situation
the longitudinal expansion makes longitudinal momenta of quarks and gluons much
smaller than their transverse momenta.

Recently, in Ref.~\cite{Romatschke:2006wg} the hard-loop effective theory for
stationary anisotropic plasmas was extended to the case of
a boost-invariant longitudinally expanding distribution of
plasma particles, the hard-expanding-loop (HEL) effective theory.
The essentially Abelian weak-field regime was worked out semi-analytically
with the result that the counterplay of increasing
anisotropy and decreasing plasma density lets Weibel instabilities grow 
exponentially in the square root of proper time, with more and more
modes becoming unstable as time goes on, but each one experiencing
a certain delay before growth kicks in.
A similar behavior was previously found in numerical studies
of initially small rapidity fluctuations in the so-called
color glass condensate framework \cite{Romatschke:2005pm,Romatschke:2006nk}.
By matching the mass scales involved with the parameters of the
saturation scenario \cite{Iancu:2003xm} the conclusion was
drawn that LHC energies will
be needed to allow for conditions where strong quark-gluon-plasma
instabilities can develop from small initial rapidity fluctuations,
leaving open however the issue of strong initial gauge fields.

In the present paper we begin the study of the evolution of 
genuinely
non-Abelian plasma instabilities in a longitudinally expanding plasma
by a lattice discretization of the HEL theory
and 1D+3V simulations. The latter have been found to give an upper
limit of the full 3+1 dimensional evolution of more generic
field configurations. The results of \cite{Bodeker:2007fw}
for strong anisotropy suggest that this upper limit may well be reached
by 3+1 dimensional plasma instabilities that start out as small
rapidity fluctuations (though not for those that are
initially non-perturbatively large). 3D+3V (as well as
2D+3V \cite{Arnold:2007tr}) real-time
lattice simulations of the HEL theory, which will be needed to address
also initially strong rapidity fluctuations,
will be the subject of follow-up work.


\section{Hard-loop effective field equations for an anisotropically expanding non-Abelian plasma}

For an ultrarelativistic plasma, 
a sufficiently small
(gauge) coupling $g$ introduces a hierarchy of scales, separating
the hard momenta $|\mathbf p|=p^0$ of plasma constituents from
the ``soft'' 
scale $\sim g\sqrt{f}\,|\mathbf p|$, where $f$ is the typical hard
particle occupation number (which may be different from order one
in strongly nonequilibrium situations). 
The soft scale is
associated with various screening phenomena
and the various branches of plasmon propagation.
Ultrasoft scales $\sim g^2f\,|\mathbf p|$ are responsible for the damping
of quasiparticles and, in or close to thermal equilibrium,
for the nonperturbative screening of chromomagnetostatic
fields.

In an anisotropic plasma, the perturbatively accessible soft scale
is also responsible for plasma instabilities, which constitute
the dominant nonequilibrium effects at weak coupling: the associated rates are
parametrically
larger than any of the scattering processes, even though the latter
are enhanced in a non-Abelian plasma.\footnote{%
As we shall see below, for strongly anisotropic plasmas the relevant
soft-scale parameters depend also importantly
on the anisotropy parameter(s) hidden in $f$.} As long as the amplitude of the
gauge fields $A\ll \sqrt{f}\,|\mathbf p|$, the evolution of the plasma instabilities
is essentially Abelian and can be studied by a perturbative linear response analysis.
For a stationary anisotropic plasma, the evolution is simply
exponential in time. When the amplitude 
becomes nonperturbatively
large, $A\gtrsim \sqrt{f}\,|\mathbf p|$, non-Abelian self-interactions
of the gauge fields become important to leading order and
require numerical evaluation, which as long
as $A\ll |\mathbf p|/g$ can be carried out consistently within
the hard-loop effective field theory framework.\footnote{For 
numerical simulations which take into account the
backreaction of the soft fields on the hard particles that come into
the play when $A\sim |\mathbf p|/g$ using
a Boltzmann-Vlasov treatment see 
Refs.~\cite{Dumitru:2005gp,Dumitru:2006pz,Dumitru:2007rp};
for numerical simulations which include backreaction using a 
statistical classical field theory
treatment see Refs.~\cite{Romatschke:2005pm,Romatschke:2006nk,Berges:2007re}.
}
In the latter, the hard particles are integrated out to produce a nonlocal and
highly nonlinear effective action which can be written in terms of a
compact integral representation \cite{Taylor:1990ia,Braaten:1992gm,Frenkel:1992ts}.
This was initially obtained for the case of thermal equilibrium
and has a straightforward generalization to the case of
stationary momentum space anisotropy \cite{Pisarski:1997cp,Mrowczynski:2004kv}.
It is of particular importance to numerical lattice studies that
the corresponding effective field equations can be made local
at the expense of introducing a continuous set of auxiliary fields
\cite{Nair:1994xs}
which arise naturally when solving gauge covariant Boltzmann-Vlasov equations
\cite{Blaizot:1994be,Blaizot:1994am,Kelly:1994dh,Blaizot:2001nr}.
In the hard loop approximation, these auxiliary fields
depend on the velocity vector of the hard particles whose
hard momentum scale is integrated out.

In Ref.~\cite{Romatschke:2006wg} this approach was extended to the
case of a nonstationary plasma with a free streaming expanding
distribution of hard particles, which we now review, filling in some
details left out in Ref.~\cite{Romatschke:2006wg}, before
proceeding with numerical real-time lattice calculations.
The latter allows us to follow the time evolution of plasma
instabilities with initially small fields into the regime where
non-Abelian self-interactions become important. The key difference
to previous hard-loop simulations of non-Abelian plasma instabilities
\cite{Rebhan:2004ur,Arnold:2005vb,Rebhan:2005re,Bodeker:2007fw}
is in the time-dependence of the
(soft-scale) parameters which determine the growth rate
of a given unstable mode and also which modes are unstable.

\subsection{Gauge-covariant Boltzmann-Vlasov equations in
a nonstationary plasma}

Assuming a color neutral background distribution function $f_0(\1p,\1x,t)$
which satisfies
\be\label{vdf0}
v\cdot \6\, f_0(\1p,\1x,t)=0,\qquad v^\mu=p^\mu/p^0,
\ee
the gauge covariant Boltzmann-Vlasov equations for colored perturbations
$\delta f_a$ of an approximately collisionless plasma have the form
\be\label{vDf}
v\cdot D\, \delta f_a(\1p,\1x,t)=g v_\mu F^{\mu\nu}_a \6^{(p)}_\nu f_0(\1p,\1x,t),
\ee
which have to be solved self-consistently with the non-Abelian Maxwell equations
\be\label{DFj}
D_\mu F^{\mu\nu}_a=j^\nu_a=
g\, t_R \int{d^3p\0(2\pi)^3} \frac{p^\mu}{2 p^0} \delta f_a(\1p,\1x,t).
\ee
Here $t_R$ is a suitably normalized group factor, while the total
number of degrees of freedom of the hard particles is taken care of
by the normalization of the distribution function $f_0$.

In a stationary 
(but possibly anisotropic) plasma $f_0$ only depends on momenta,
and (\ref{vdf0}) is satisfied trivially. Here we 
shall consider the generalization
to a plasma which expands longitudinally, which should be a good approximation
for the initial stage of a parton gas produced in a heavy ion collision
as long as the transverse dimension of the system is sufficiently large.
Assuming furthermore boost invariance in rapidity
\cite{Bjorken:1983qr}
and isotropy in the transverse directions,
the unperturbed
distribution function $f_0$, being a Lorentz scalar, has the form
\cite{Baym:1984np,Mueller:1999pi}
\be
f_0(\mathbf p,x)=f_0(p_\perp,p^z,z,t)=f_0(p_\perp,p'^z,\tau)
\ee
where the transformed longitudinal momentum is
\be
p'^z=\gamma(p^z-\beta p^0),\quad
\beta=z/t,\quad \gamma=t/\tau,\quad \tau=\sqrt{t^2-z^2},
\ee
with $p^0=\sqrt{p_\perp^2+(p^z)^2}$ for ultrarelativistic (massless) particles.

\subsection{Comoving coordinates}

It is convenient to switch to comoving coordinates
\bea
t=\tau \cosh\eta,\quad &&\beta=\tanh\eta,\nn\\
z=\tau \sinh\eta,\quad  &&\gamma=\cosh\eta,
\eea
i.e.\ a coordinate system with metric
$ds^2=d\tau^2-d\mathbf x_\perp^2-\tau^2 d\eta^2$.
We introduce the notation $\tilde x^\alpha=(x^\tau,x^{\9{i}},x^\eta)=
(\tau,x^1,x^2,\eta)$
with indices from the beginning of the Greek alphabet
for these new coordinates.
Note that in the latter the
indices $i,j,\ldots$ are restricted to the two transverse spatial
coordinates.

In what follows we shall 
not deal with space-time covariant derivatives and Christoffel
symbols, but write everything in terms of explicit derivatives.
In particular the gauge covariant derivative always 
means\footnote{Recall that
$A^\mu=(\phi,\vec A)$ with 4-index up. 
Thus $\tilde A_\alpha=(A_\tau,-A^x,-A^y,A_\eta)$.} 
$\tilde D_\alpha=\tilde\6_\alpha-ig[\tilde A_\alpha,\cdot]$.
Being a two form (where indices are naturally down),
the field strength retains its usual form:
$\tilde F_{\alpha\beta}=\tilde\6_\alpha \tilde A_\beta
-\tilde\6_\beta \tilde A_\alpha
-ig[\tilde A_\alpha,\tilde A_\beta]$.
The (non-Abelian) Maxwell equations do involve additional
terms, but they can be written compactly as
\be
{1\0\tau}\tilde D_\alpha(\tau \tilde F^{\alpha\beta}) \equiv
{1\0\tau}\tilde D_\alpha\left[\tau g^{\alpha\gamma}(\tau) g^{\beta\delta}(\tau)
\tilde F_{\gamma\delta}\right]=\tilde j^\beta.
\ee

In addition to space-time rapidity $\eta$, we also introduce
momentum space rapidity $y$ for the massless particles according to
\be
p^\mu=p_\perp(\cosh y,\cos\phi,\sin\phi,\sinh y).
\ee
In comoving (tilde) coordinates, we then have
\bea
\tilde 
p^\tau&=&\sqrt{p_\perp^2+\tau^2(\tilde p^\eta)^2}\nn\\
&=&\cosh\eta\, p^0-\sinh\eta\, p^z
=p_\perp \cosh(y-\eta),\\
\tilde
p^\eta&=&-\tilde p_\eta/\tau^2=(\cosh\eta\, p^z-\sinh\eta\, p^0)/\tau\nn\\
&=&p'^z/\tau
=p_\perp \sinh(y-\eta)/\tau.
\eea

Instead of the light-like vector $v^\mu=p^\mu/p^0$ 
containing a unit 3-vector that was
used in Eqs.~(\ref{vdf0}) and (\ref{vDf}), we shall
define the
new quantity 
\be
\tilde V^\alpha = {\tilde p^\alpha \0 p_\perp} =
\left(\cosh (y-\eta),\,\cos\phi,\,\sin\phi,\,{1\0\tau}\sinh (y-\eta)\right),
\label{velocityDef1}
\ee
which is normalized so that it has a unit 2-vector in the transverse plane.

\subsection{Longitudinally expanding free streaming background solution}


Eq.~(\ref{vdf0}), involving space-time derivatives at fixed
$\mathbf p_\perp$ and $p^z$, can be rewritten as
\be
(\tilde p\cdot \tilde \6) f_0\Big|_{y,\1p_\perp}=0.
\ee
Because
\bea
\tilde p^\tau \6_\tau \tilde p_\eta(\tilde x)\Big|_{y,\1p_\perp}
&=&-p_\perp^2 \sinh(y-\eta)\cosh(y-\eta)\nn\\
&=&-\tilde p^\eta\6_\eta \tilde p_\eta(\tilde x)\Big|_{y,\1p_\perp}
\label{petaconstancy}
\eea
this can be solved by $f_0(\1p,\1x,t)=f_0(\1p_\perp,\tilde p_\eta(x))
=f_0(\1p_\perp,-p'^z(x)\tau(x))$.

In the following we shall use\footnote{Notice 
that it would be straightforward to relax the assumption
of momentum-space isotropy in the transverse directions.}
\be\label{faniso}
f_0(\mathbf p,x)=f_{\rm iso}\left(\sqrt{p_\perp^2+({p'^z\tau\0\tau_{\rm iso}})^2}\right)
=f_{\rm iso}\left(\sqrt{p_\perp^2+\tilde p_\eta^2/\tau_{\rm iso}^2}\right)
\ee
which corresponds to local isotropy on the hypersurface $\tau=\tau_{\rm iso}$,
and increasingly oblate momentum space anisotropy at $\tau>\tau_{\rm iso}$
(but prolate anisotropy for $\tau<\tau_{\rm iso}$).
Since a plasma description does not make sense at arbitrarily small
times and so time evolution will have to start at a nonzero proper time $\tau_0$,
the time $\tau_{\rm iso}$ may be entirely fictitious
in the sense of pertaining to the pre-plasma 
(glasma \cite{Lappi:2006fp,Romatschke:2006nk}) phase.
This will in fact
be the case in the numerical simulations below, where we shall start already
with oblate anisotropy by choosing $\tau_{\rm iso} < \tau_0$.

In a comoving frame,
the energy density and pressure components of the hard particle background
can be determined by evaluating $T^{\alpha\beta}_{\rm part.}=
(2\pi)^{-3}\int d^2p_\perp dy\, \tilde p^\alpha \tilde p^\beta f_0$, 
which yields
\begin{eqnarray}
\mathcal E_{\rm part.}(\tau) &=& T^{\tau\tau}_{\rm part.} = \frac1{2}
\left[\frac1{\bt^2}+\frac{\arcsin\sqrt{1-\bt^{-2}}}{\sqrt{\bt^2-1}}\right]\mathcal E_{\rm iso} ,
\\
P_T^{\rm part.}(\tau)&=&\frac12 T^{\9i \9i }_{\rm part.}=
\frac1{4(\bt^2-1)}
\left[1+\frac{\bt^2-2}{\sqrt{\bt^2-1}}\arcsin\sqrt{1-\bt^{-2}} \right]\mathcal E_{\rm iso},
\label{particlePT}
\\
P_L^{\rm part.}(\tau)&=&-T^\eta_{{\rm part.}\eta} = \frac1{2(\bt^2-1)}
\left[-\frac1{\bt^2} +  \frac{\arcsin\sqrt{1-\bt^{-2}}}{\sqrt{\bt^2-1}} \right]\mathcal E_{\rm iso},
\label{particlePL}
\end{eqnarray}
where 
$\mathcal E_{\rm iso}=\mathcal E_{\rm part.}(\tau_{\rm iso})$,
$\bt \equiv \tau/\tau_{\rm iso}$ and we have assumed $\bt\ge1$.
For $\bt\gg1$ we have
\be
P^{\rm part.}_T \to \frac\pi8 \mathcal E_{\rm iso}\bt^{-1},\qquad
P^{\rm part.}_L \to \frac\pi4 \mathcal E_{\rm iso}\bt^{-3}.
\ee
The energy density follows from 
$\mathcal E_{\rm part.}\equiv 2P^{\rm part.}_T+P^{\rm part.}_L$.

The particle distribution function (\ref{faniso}) has the same form as the one used in 
Refs.~\cite{Romatschke:2003ms,Romatschke:2004jh,Rebhan:2004ur,Rebhan:2005re},
but the anisotropy parameter $\xi$ therein\footnote{The anisotropy parameter
$\theta$ used in Ref.~\cite{Arnold:2007cg} 
is related to $\xi$ by $\xi\sim\theta^{-2}$.}
is now space-time dependent
according to 
\be\label{xi}
\xi(\tau)=(\tau/\tau_{\rm iso})^2-1,
\ee
and the
normalization factor $N(\xi)$ of 
Ref.~\cite{Romatschke:2004jh,Rebhan:2004ur,Rebhan:2005re}
is unity.

The behavior $\xi\sim\tau^2$ at large $\tau$ is a consequence of
having a free-streaming background distribution. In a more realistic
collisional plasma, $\xi$ will have to grow slower than this.
In the first stage of the original bottom-up scenario \cite{Mueller:2005un}, ignoring
plasma instabilities, one would have had $\xi\sim\tau^{2/3}$. 
In Ref.~\cite{Bodeker:2005nv} it was argued that plasma instabilities
reduce the exponent to $\xi\sim\tau^{1/2}$, whereas Ref.~\cite{Arnold:2007cg}
recently presented arguments in favor of $\xi\sim\tau^{1/4}$.
All these scenarios have $\xi\gg1$, so below we shall concentrate
on the case $\tau_{\rm iso}<\tau_0$ and thus high anisotropy for all $\tau>\tau_0$,
but in the idealized case of a collisionless free-streaming expansion.

\subsection{HEL effective field equations}

Transforming the gauge-covariant Vlasov equation to comoving
coordinates one can write
\be\label{VDf}
\tilde V\cdot \tilde D\, \delta f^a\big|_{p^\mu}=g \tilde V^\alpha 
\tilde F_{\alpha\beta}^a \tilde \6_{(p)}^\beta f_0(\1p_\perp,\tilde p_\eta),
\ee
where the derivative on the left-hand side has to be taken at fixed $p^\mu$
as opposed to fixed $\tilde p^\alpha$. On the right-hand side the derivative
with respect to momenta is at fixed $x$, but the transformation from
$x$ to $\tilde x$ does not depend on momenta anyway. However, in the
following it will be important to write the right-hand side in terms
of $\tilde \6_{(p)}^\beta f_0(\1p_\perp,\tilde p_\eta)$ with index up so that
this factor depends only on $\1p_\perp$ and $\tilde p_\eta$ and not
additionally on $\tau$. This means in particular that
$p\cdot\6 \,(\tilde \6_{(p)}^\beta f_0)|_p=
\tilde p\cdot\tilde\6 \,(\tilde \6_{(p)}^\beta f_0)|_p=0$.

Eq.~(\ref{VDf}) can then be solved in terms of an auxiliary field
$\tilde W_\beta(\tilde x;\phi,y)$ which satisfies
\be\label{VDW}
\tilde V\cdot \tilde D\, \tilde W_\beta\big|_{\phi,y}=\tilde V^\alpha \tilde F_{\beta\alpha} \, ,
\ee
and
\be\label{Wdef}
\delta f(x;p)=-g \tilde W_\beta(\tilde x;\phi,y) 
\tilde\6_{(p)}^\beta f_0(p_\perp,\tilde p_\eta).
\ee

The field $\tilde W_\beta(\tilde x;\phi,y)$ is indeed analogous
to the auxiliary field $W_\nu(x;\1v)$ of the (static) hard-loop formalism
\cite{Mrowczynski:2004kv} because for a given space-time point
it only depends on the 3-velocity of the hard particles,
$\1v=(\cos\phi,\sin\phi,\sinh y)/\cosh y$, and not on their
energy $p^0$. Notice that only with index down its equation of
motion (\ref{VDW}) is formally the same as in the static situation.

Expressed in terms of the auxiliary field $\tilde W$, 
the induced current in comoving
coordinates reads
\bea\label{tjind}
\tilde j^\alpha[A] &=& - \frac{g^2 t_R}{2}
\int {d^3p\over(2\pi)^3} 
{1\over p^0} \,\tilde p^\alpha\, {\partial f_0(p_\perp,\tilde p_\eta) 
\over \partial \tilde p_\beta}
\tilde W_\beta(\tilde x;\phi,y)\nn\\
&=&+g^2t_R
\int {d^2p_\perp\,dp_\eta\over(2\pi)^3} 
{1\over2\tau p^\tau} \,\tilde p^\alpha\, {\partial f_0(p_\perp,\tilde p_\eta) 
\over \partial \tilde p_\beta}
\tilde W_\beta 
\nn\\
&=&-g^2t_R
\int_0^\infty {p_\perp dp_\perp \over 8\pi^2}
\int_0^{2\pi} {d\phi \0 2\pi}
\int_{-\infty}^\infty dy \,
\tilde p^\alpha\, {\partial f_0 
\over \partial \tilde p_\beta}
\tilde W_\beta 
\nn\\
\eea
where for each $(\phi,y)$ (i.e., fixed $\1v$) the scale
$p_\perp$ (related to energy by $p^0=p_\perp\cosh y$) can be integrated out.

With a distribution function that is
even in $\1p_\perp$ and $\tilde p_\eta$ as in (\ref{faniso}),
covariant current conservation can be verified without having
to integrate partially with respect to $p$. (This proves to be
helpful for the lattice discretization below, where all integrals
will be replaced by discrete sums.)
The current $j^\mu$
in ordinary coordinates is given by Eq.~(\ref{tjind}) by dropping
the tilde on $\tilde j^\alpha$ and $\tilde p^\alpha$ only. 
Starting from the first line of (\ref{tjind}), we can then use
$D\cdot p=p\cdot D=\tilde p\cdot \tilde D$, and  
$(p\cdot\6){\6f_0/\6\tilde p_\beta}|_p=0$ and finally
(\ref{VDW}) (with $\tilde V$ replaced by $\tilde p^\alpha$).
Changing the integration variables to $\tilde p$ like in the
second line of (\ref{tjind}) we obtain
\be\label{Dj}
D\cdot j=g^2 t_R\int {d^2p_\perp\,dp_\eta\over(2\pi)^3} 
{1\over2\tau \sqrt{p_\perp^2+p_\eta^2/\tau^2}}\nn\\
\left[ 
{\6f_0\0\6p^{\9{i}}}\tilde F^{\9{i}}{}_\gamma \tilde p^\gamma
+{\6f_0\0\6p_\eta}\tilde F_{\eta\gamma}\tilde p^\gamma \right].
\ee
This vanishes already by symmetry when ${\6f_0/\6p^{\9{i}}}$ and
${\6f_0/\6p_\eta}$ are odd functions in $p^{\9{i}}$ and
$p_\eta$, respectively.

Specializing to the background distribution function (\ref{faniso})
we have
\bea
\tilde\partial_{(p)}^\beta f_0&=&f_0'\tilde\partial_{(p)}^\beta
\sqrt{p_\perp^2+\tilde p_\eta^2/\tau_{\rm iso}^2}\nn\\
&=&
{\left(0,-\cos\phi,-\sin\phi,-{\tau\0\tau_{\rm iso}^2}\sinh(y-\eta)\right)
\0\sqrt{1+{\tau^2\0\tau_{\rm iso}^2}\sinh^2(y-\eta)}} \, , 
\eea
and we get
\be
\tilde j^\alpha=-m_D^2\, \2\int_0^{2\pi} {d\phi \0 2\pi}
\int_{-\infty}^\infty dy \, 
\tilde V^\alpha
\left(1+{\tau^2\0\tau_{\rm iso}^2}\sinh^2(y-\eta)\right)^{-2}
\mathcal W(\tilde x;\phi,y) \, ,
\label{current}
\ee
where
\bea
&&\mathcal W = \tilde V^{\9{i}} W_{\9{i}}
-{1\0\tau_{\rm iso}^2}\tilde V_\eta\, \tilde W_\eta,\qquad \nn\\ &&
\tilde V^{\9{i}}=(\cos\phi,\sin\phi) \, ,\quad
\tilde V_\eta = -\tau \sinh(y-\eta) \, ,
\eea
and
\be
m^2_D=-g^2t_R \int_0^\infty {dp\,p^2\0(2\pi)^2} f'_{\rm iso}(p) \, .
\ee
The mass parameter $m_D$ equals the Debye mass at the (possibly fictitious
because pre-plasma)
time $\tau_{\rm iso}$.

Because $\tilde V\cdot \tilde D$ commutes with the coefficients
of $\tilde W_{\9{i}}$ and $\tilde W_\eta$ 
appearing in the definition of $\mathcal W$ (in particular
$[\tilde V\cdot \tilde D,\tilde V_\eta]=0$, cf.~(\ref{petaconstancy})),
we do not need to evolve the components $\tilde W_\beta$ separately,
but only the combination $\mathcal W$, which is governed by
\be\label{VDtildeW}
\tilde V\cdot \tilde D\, \mathcal W
= \left(\tilde V^{\9{i}} \tilde F_{\9{i}\tau}+{\tau^2\0\tau_{\rm iso}^2}\tilde V^\eta \tilde F_{\eta\tau}\right)\tilde V^\tau 
+\tilde V^{\9{i}}\tilde V^\eta \tilde F_{\9{i}\eta}\left(1-{\tau^2\0\tau_{\rm iso}^2}\right).
\ee
For $\tau=\tau_{\rm iso}$ only $F_{\alpha\tau}$ (the 
electric field components in the
comoving frame\footnote{From here on we shall drop the tilde on the
quantities in the comoving frame, which will be used exclusively in what
follows.}) appear on the r.h.s., whereas for $\tau\not=\tau_{\rm iso}$
magnetic fields come into the play, opening the door for magnetic
instabilities.

This single equation for $\mathcal W$ together with the Yang-Mills equations
and the algebraic relation between $j$ and $\mathcal W$ 
closes our equations of motion. To solve them, we adopt the comoving
temporal gauge $A^\tau=0$ and introduce canonical conjugate field momenta
for the remaining gauge fields according to
\be
\Pi^{\9{i}} = \tau \6_\tau A_{\9{i}} = 
-\tau \6_\tau A^{\9{i}}=-\Pi_{\9{i}} \, ,
\ee
and
\be
\Pi^\eta={1\0\tau}\6_\tau A_\eta \, .
\ee
Notice that transverse (comoving) electric field components 
differ from $\Pi^{\9{i}}$ by a factor of $\tau$:
\be
E^{\9{i}}=\tau^{-1} \Pi^{\9{i}} \, .
\ee
In contrast to most of the literature on the color glass condensate
framework, we shall reserve the symbol $E$ for
the electric field and denote the canonical conjugate field momenta
by $\Pi$.

In terms of fields and conjugate momenta,
the Yang-Mills equations take the form
\bea
\tau \6_\tau \Pi^\eta&=&j_\eta-D_{\9{i}} F^{\9{i}}{}_\eta\,,\\
\tau^{-1}\6_\tau \Pi_{\9{i}}&=&j^{\9{i}}
-D_{\9{j}}F^{{\9{j}}{\9{i}}}
-D_\eta F^{\eta{\9{i}}}\,.
\eea

\subsection{1D+3V equations}

A linear response analysis (appropriate for small gauge field amplitudes)
shows that the most unstable modes of an anisotropic plasma are those
whose wave vector is oriented along the direction of anisotropy.

We therefore begin by
considering only initial conditions and thus
solutions which are constant in the transverse
directions (i.e., neglecting transverse dynamics), 
$\partial_{\9{i}} A^\alpha\equiv 0$. Hence,
$D^{\9{i}}=-ig[A^{\9{i}},\cdot]$ and the Yang-Mills equations
reduce to that of a 1+1 dimensional theory with $A^{\9{i}}$
acting as adjoint scalars.

We then have
\bea\label{Pieom1+1}
{1\0\tau}\6_\tau \Pi_{\9{i}}&=&j^{\9{i}}
+g^2 i[A^{\9{j}},i[A^{\9{j}},A^{\9{i}}]]
+{1\0\tau^2}D_\eta^2 A^{\9{i}} \, , \\
\tau\6_\tau \Pi^\eta&=&j_\eta+ig[A^{\9{i}},D_\eta A^{\9{i}}] \, ,
\eea
as dynamical Yang-Mills equations, and
\be
\tau j^\tau = D_\eta \Pi^\eta-ig[A^{\9{i}},\Pi_{\9{i}}] \, ,
\label{gausslawcontinuum}
\ee
as Gauss law constraint.

The current $j^\alpha$ 
is a linear functional of $\mathcal W$, given by Eq.~(\ref{current})
as before, but
the equation of motion for $\mathcal W$, Eq.~(\ref{VDtildeW}),
reduces to
\bea\label{DVW1+1}
\6_\tau \mathcal W(\tau,\eta;\phi,y) &=& 
{\tH\0\tau}D_\eta \left(\left(1-{\tau^2\0\tau_{\rm iso}^2}\right)v^{\9{i}} A^{\9{i}}
-\mathcal W \right)\nn\\
&&- {ig\0\cH}[v^{\9{i}} A^{\9{i}},\mathcal W]
+{1\0\tau}v^{\9{i}}\Pi_{\9{i}}-{\tau^2 \sH \0 \tau_{\rm iso}^2} \Pi^\eta \, .
\eea
All fields here depend on the two remaining space-time variables
$\tau,\eta$, and the auxiliary adjoint-scalar
field $\mathcal W$ additionally depends
on the momentum space variables $\phi,y$ which parametrize
the 3-velocity in the colored fluctuations $\delta f^a$, cf.\
Eq.~(\ref{Wdef}).

In the present paper we shall restrict our attention to this
dimensionally reduced situation, which in conventional plasma
literature would be referred to as 1D+3V, postponing the study
of the more general 2D+3V and 3D+3V cases to future publications.

\section{Lattice discretization and numerical results}

\subsection{Methods}

For a numerical evaluation of Eqs.~(\ref{Pieom1+1})--(\ref{DVW1+1}) 
together with Eq.~(\ref{current}) we discretize proper time starting 
with finite $\tau_0>0$ and time step $\epsilon$. The space-time 
rapidity coordinate $\eta$ is made periodic and discrete with 
$N_\eta$ points and
(dimensionless) spacing $a$ covering a rapidity interval
$(-N_\eta a/2,N_\eta a/2)$. 
The (matrix-valued) fields $A^x$, $A^y$, 
and $\mathcal W_{\phi,y}$ are defined on the sites of the 1-dimensional 
rapidity lattice, while the conjugate momenta $\Pi_x$, 
$\Pi_y$, and $\Pi^\eta$ are defined on the temporal links. The gauge 
field $A_\eta$ is replaced by the spatial link variable $U=\exp 
igaA_\eta$.

The integration over the momentum-space variables $\phi$ and $y$ in 
Eq.~(\ref{current}) has to be discretized such that covariant current 
conservation is preserved manifestly. When expressed in terms of 
$\phi$ and $y$ integrals, the integrand in (\ref{Dj}) is either odd in 
$y-\eta$ or multiplied by $\sin\phi$ or $\cos\phi$. In order that 
discretization of $y$ and $\phi$ respect manifest covariant current 
conservation, we thus need to respect reflection invariance in $\phi$ 
and $y-\eta$. The angular variable is made discrete with uniform 
spacing $2\pi/N_\phi$, but for $\bar y\equiv y-\eta$ we shall consider 
two possibilities. In method A we shall discretize the interval $-
\Lambda_y \le \bar y \le \Lambda_y$ uniformly with spacing $2 
\Lambda_y/(N_y+1)$, and in method B we make the substitution $\bar 
y=\atanh x$ and discretize the range $-1 + \Delta x \le x \le 1 - \Delta 
x$ with uniform spacing $\Delta x = 1/N_x$. Because of the $\eta$ 
dependence of the shifted variable $\bar y$, the lattice equation of 
motion for the auxiliary fields $\mathcal W_{\phi,\bar y}$ that live 
on the $\bar y$ boundary have to be completed by boundary conditions 
for $\mathcal W$ in the $\bar y$ variable. For the $\mathcal W$ fields 
we do not impose periodicity, but instead take the Neumann condition 
${\6\mathcal W/\6\bar y}=0$ at the $\bar y$ boundary.

\begin{figure}
\centerline
{\includegraphics[width=0.55\linewidth]{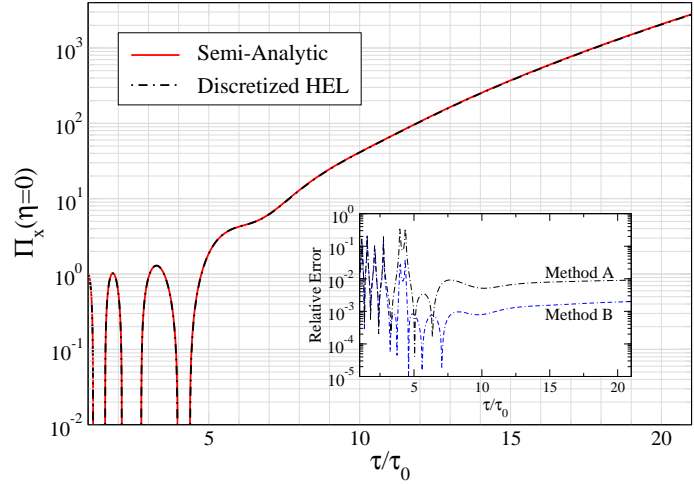}}
\caption{\label{fig1}
Proper-time evolution of the canonical field momentum $\Pi_x(\eta=0)$ of a 
single Abelian mode with rapidity wave number
$\nu = 10.053$. Solid line is the semi-analytic 
result of Ref.~\cite{Romatschke:2006wg} and the dot-dashed lines 
are the results obtained from our 1+1 numerical solutions using two 
different methods (A and B) for discretizing the shifted momentum-space 
rapidity $\bar y=y-\eta$. Inset shows 
relative error of the two methods. Run was made
using $\tau_{\rm iso}=0.1$, $\tau_0=1.0$, $m_D = 10$, $a = 0.0025$,
$\epsilon = 0.001$, $N_\eta=250$, $N_\phi=8$, and 
(Method A) $N_y=1000$, (Method B) $N_x=1000$.
}
\end{figure}

In Fig.~\ref{fig1} we show the evolution of the conjugate momentum 
$\Pi_x\propto\cos(\nu\eta)$ 
in the case that the gauge group is taken to be Abelian U(1). 
The system is initialized with a single Abelian U(1) mode with only 
$\Pi_x$ initialized with 
rapidity wave number $\nu=16\pi/5=10.053\ldots$ in order to 
facilitate comparisons with semi-analytic results obtained in an 
earlier work \cite{Romatschke:2006wg} where, for the Abelian case, the 
equations of motion for the $\mathcal W$ field have been solved in 
terms of integro-differential equations. Fig.~\ref{fig1} compares a 
semi-analytic result obtained from the latter with results obtained 
using the two different methods of discretization described above and 
detailed in Apps.~\ref{methodAapp} and \ref{methodBapp}.  As can be 
seen from this figure both numerical discretizations reliably 
reproduce the Abelian U(1) semi-analytic result.  In the inset we 
compare the relative error defined as the difference of the time 
evolution obtained from methods A or B with the semi-analytic result 
over the sum (relative percentage error).  As can be seen from this 
inset method B seems to perform better at late times so unless 
otherwise indicated all final results presented will be using method 
B. However, in practice, we have made runs comparing the predictions 
of methods A and B in all cases and find that there is very little 
difference between the results obtained with the two methods.

\subsection{Single-mode results}

\begin{figure}
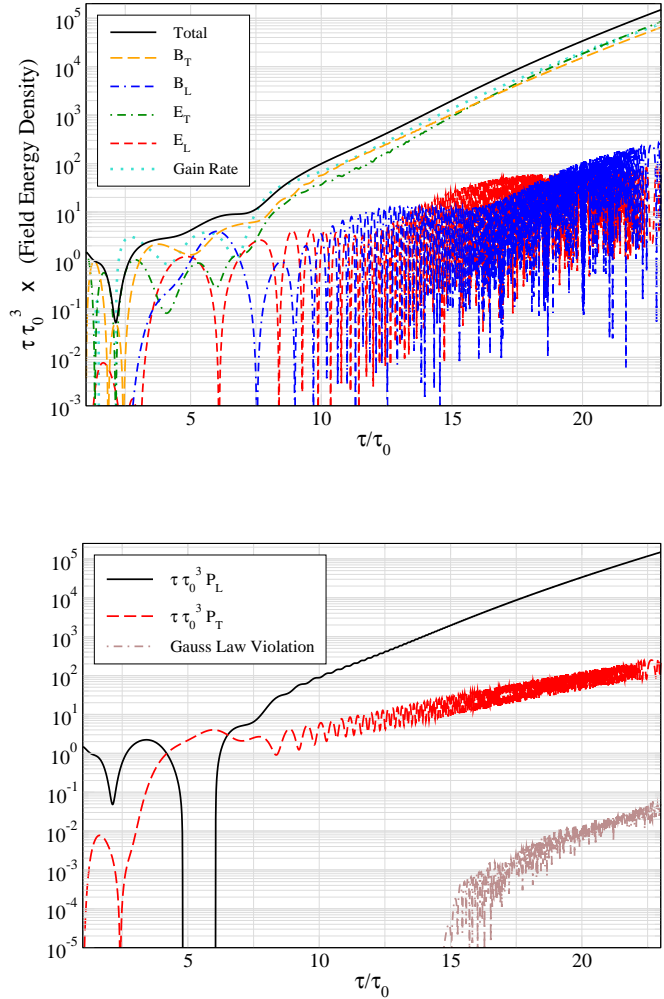

\centerline
{\includegraphics[width=0.53\linewidth]{fig2-a.eps}~~~~}
\vspace{1.1cm}
\centerline
{\includegraphics[width=0.5\linewidth]{fig2-b.eps}}
\caption{\label{fig2}
Results from a run with a single non-Abelian mode
with $\nu = 10.053$.  In the top panel (a) we show the proper-time dependence
of the chromo-field energy densities.  In the lower panel (b) we show the 
longitudinal and transverse pressures along with our numerical 
Gauss law violation.  
Run was made using 
$\tau_{\rm iso}=0.1$, $\tau_0=1.0$, $m_D = 10$, $a = 0.0025$,
$\epsilon = 0.001$, $N_\eta=500$, $N_x=100$, and $N_\phi=100$.
}
\end{figure}

In Fig.~\ref{fig2} we show results of a simulation of a single SU(2) 
mode with rapidity wave number $\nu=10.053$ 
(same mode as Fig.~\ref{fig1} but now also
with the color direction
rotating with period $2\pi/\nu$ in space-time rapidity $\eta$).  In 
Fig.~\ref{fig2}a we show the proper-time evolution of the magnetic, 
electric, and total field energy densities in units where $\tau_0=1$ 
and, following Ref.~\cite{Romatschke:2006nk}, 
scaled with a factor of $\tau$. Because the energy in the hard 
particles is dropping proportional to $\tau^{-1}$, this corresponds
to giving the various soft energy densities in terms of the hard energy
density (times a parametrically small number $\sim g^2$
since the hard energy density is 
assumed to be much larger than the soft ones
in order that the hard-loop approximation be applicable.)

The various components of the (soft) field energy density are defined by
\bea
\mathcal E=\mathcal E_T+\mathcal E_L&=&
\mathcal E_{B_T}+\mathcal E_{E_T}+\mathcal E_{B_L}+\mathcal E_{E_L}\nn\\
&=&\tr\left[ \tau^{-2} F_{\eta \9i}^2 +
\tau^{-2}\Pi_{\9i}^2 +
F_{xy}^2+ \left(\Pi^{\eta}\right)^2 \right].
\eea
Because of the expansion of the system,
the total energy density $\mathcal E$ is not conserved, even when the
induced current (\ref{current}) is identically zero. In this case
the time dependence is governed by the fact that the Hamiltonian
density \cite{Romatschke:2006nk} $\mathcal H=\tau\mathcal E$ satisfies 
\be
\frac{d}{d\tau}\mathcal H=\frac{\6}{\6\tau}\mathcal H
=\mathcal E_{L}-\mathcal E_T \, ,
\ee
and therefore
\be
\frac{d}{d\tau}\mathcal E|_{j\equiv0}=-\frac{2}{\tau}\mathcal E_T|_{j\equiv0}\,.
\ee

In the presence of a plasma of hard particles and thus nonvanishing
induced current $j$ we define the net energy gain rate by
\be\label{REG}
R_{\,\rm Energy\;Gain} \; \equiv \; 
  \frac{d{\cal E}}{d\tau} + \frac{2}{\tau}{\cal E}_T\,,
\ee
which in the plots showing the energy densities is included as the dotted line
marked ``Gain Rate''. The latter gives the rate of energy transfer
from the free-streaming hard 
particles into the collective chromo-fields. 
As can be seen from Fig.~\ref{fig2} for SU(2) 
the single mode evolution is quite complicated with all field components 
being dynamically generated; however, at late times transverse 
chromoelectric and chromomagnetic fields exponentially dominate.

In Fig.~\ref{fig2}b we plot the longitudinal and transverse field 
pressures generated during the system's dynamical evolution.  These 
are obtained from \cite{Romatschke:2006nk}
\bea
P_L &=& {\cal E}_T - {\cal E}_L \, , \nonumber \\
P_T &=& {\cal E}_L \, ,
\eea
where as before ${\cal E}_T$ is the sum of the energy density coming 
from transverse electric and magnetic fields and  ${\cal E}_L$ is the 
sum of the energy density coming from longitudinal electric and 
magnetic fields. As shown in Fig.~\ref{fig2}b the system
generates both longitudinal and transverse pressures.  At short times 
($\tau/\tau_0 \sim$ 5-6) for this single mode evolution we find that 
the longitudinal pressure becomes momentarily negative; however, at 
late times the effect of the chromo-field instability is to generate 
exponentially large longitudinal field pressure, whereas the
longitudinal pressure of the (free-streaming) particles
drops according to $\tau P_L^{\rm part.}\sim\tau^{-2}$.

\begin{figure}
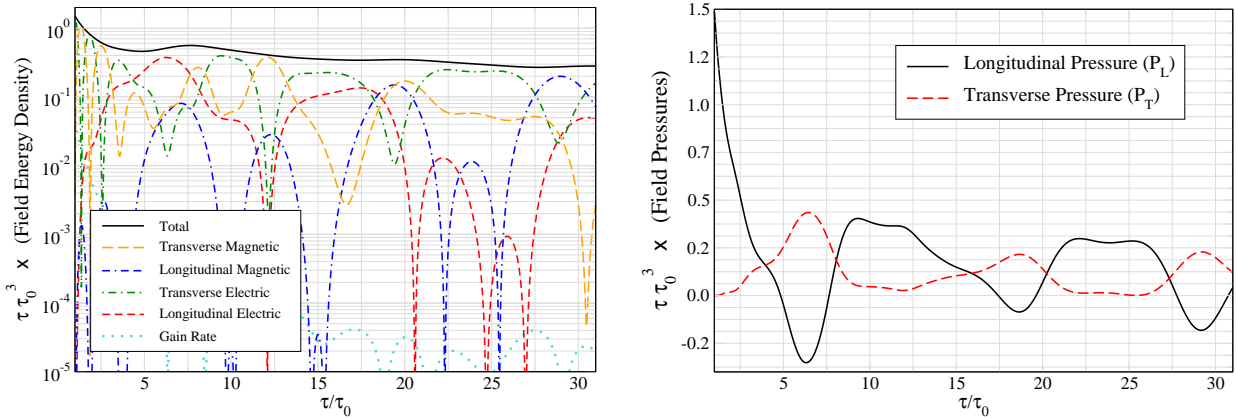

\centerline
{\includegraphics[width=0.48\linewidth]{fig3-a.eps}
\hspace{0.02\linewidth}
\includegraphics[width=0.48\linewidth]{fig3-b.eps}}
\caption{\label{fig3}
Results from a run with a single non-Abelian mode
with rapidity wave number
$\nu = 10.053$ in which we have decoupled the hard particle currents
($j=0$) so that we are simply solving the Yang-Mills equations in the expanding
metric.  In the left panel (a) we show the proper-time dependence
of the chromo-field energy densities.  In the right panel (b) we show the 
longitudinal and transverse pressures.  Gauss law is obeyed exactly
by our algorithm in this case.  Run was made
using $\tau_{\rm iso}=0.1$, $\tau_0=1.0$, $m_D = 10$, $a = 0.0025$,
$\epsilon = 0.001$, and $N_\eta=500$.
}
\end{figure}

\begin{figure}
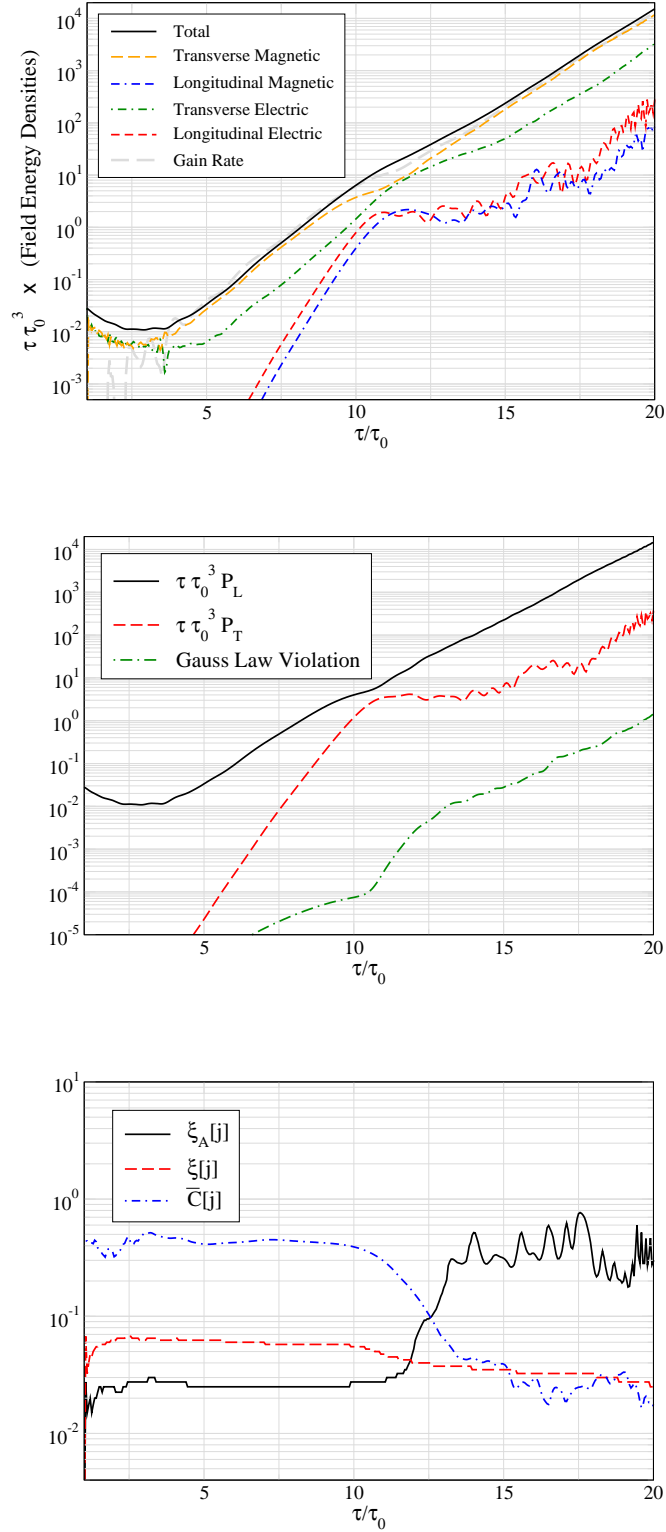

\centerline
{\includegraphics[width=0.53\linewidth]{fig4-a.eps}~~~~}
\vspace{1.1cm}
\centerline
{\includegraphics[width=0.5\linewidth]{fig4-b.eps}}
\vspace{1.1cm}
\centerline
{\includegraphics[width=0.5\linewidth]{fig4-c.eps}}
\caption{\label{fig4}
Results from non-Abelian run initialized with a random superposition 
of discrete electric modes (cutoff white noise).  In the top panel (a) 
we show the proper-time dependence of the chromo-field energy 
densities and the energy gain rate (\ref{REG}) times an extra factor of
$\tau_0$.  In the middle panel (b) we show the longitudinal and 
transverse pressures along with our numerical Gauss law violation.  In 
bottom panel (c) we show the correlations $\xi_A[j]$, $\xi[j]$, and 
${\bar C}[j]$.  Run was made using $\tau_{\rm iso}=0.1$, $\tau_0=1.0$, 
$m_D = 10$, $\sigma=0.03$, $\Lambda_\nu=20$, $a = 0.0025$, $\epsilon = 
0.00025$, $N_\eta=1000$, $N_x=100$, and $N_\phi=100$.
}
\end{figure}

Also shown in Fig.~\ref{fig2}b is our measure of violation of Gauss 
law which is determined by evaluating the $\tau$-component of the 
equations of motion as detailed in Eqs.~(\ref{gausslawcontinuum}) and 
(\ref{gausslawlattice}).  As can be seen from this figure although our 
violation of the Gauss law constraint grows with time, it is 
numerically under control and always orders of magnitude below the 
field energy density.  The amount of violation can be systematically 
reduced by taking finer lattices in $\eta$ and velocity space.  We 
have found that our results for the time evolution of the energy 
densities, pressures, etc.\ remain the same as our numerical Gauss law 
violation is reduced giving us confidence in our algorithm.  As a 
general rule we have always terminated our runs when the Gauss law 
violation becomes of order one.

For comparison in Fig.~\ref{fig3} we show the evolution of the field 
energy densities in the case of pure Yang-Mills evolution.  This is 
obtained by decoupling the free-streaming particle currents by setting 
$j^\alpha$ to zero in the field equations of motion.  From 
Fig.~\ref{fig3}a we see that in the case of pure Yang-Mills evolution 
the field energy density decreases over the entire time interval 
shown. The ``Gain Rate'' control variable is
approximately zero and shows the level of discretization errors.
In addition we see that although both longitudinal and 
transverse pressures are generated they are of much smaller magnitude 
than those generated when the free-streaming particle currents are 
coupled into the Yang-Mills equations. Therefore, we have demonstrated 
that coupling in the particle currents generates qualitatively 
different field dynamics.

\begin{figure}
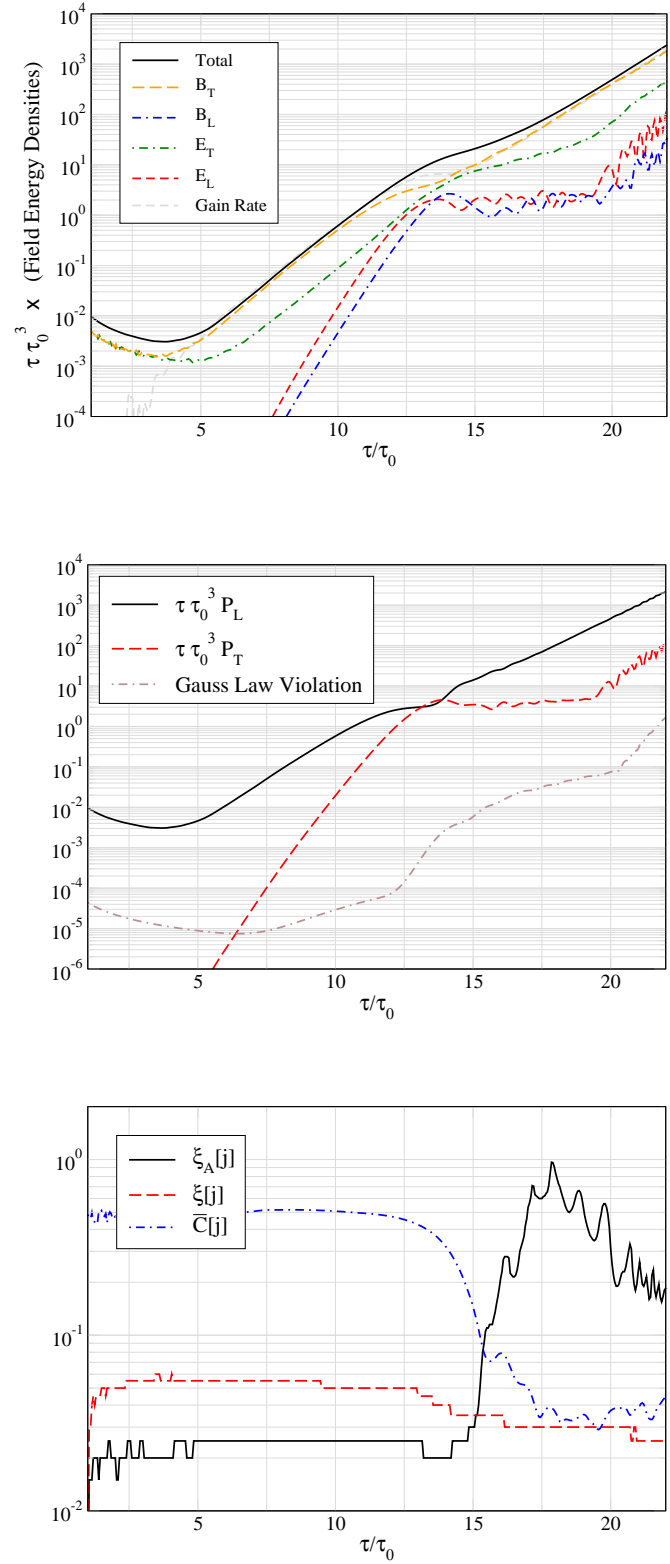

\centerline
{\includegraphics[width=0.53\linewidth]{fig5-a.eps}~~~~}
\vspace{1.1cm}
\centerline
{\includegraphics[width=0.5\linewidth]{fig5-b.eps}}
\vspace{1.1cm}
\centerline
{\includegraphics[width=0.5\linewidth]{fig5-c.eps}}
\caption{\label{fig5}
Results from non-Abelian run initialized with FGM initial conditions.  In the top panel (a) we show the proper-time dependence
of the chromo-field energy densities and the energy gain rate (\ref{REG}) times an extra factor of
$\tau_0$. In the middle panel (b) we show the 
longitudinal and transverse pressures along with our numerical 
Gauss law violation.  In bottom panel (c) we show the correlations
$\xi_A[j]$, $\xi[j]$, and ${\bar C}[j]$.  Run was made
using $\tau_{\rm iso}=0.1$, $\tau_0=1.0$, $m_D = 10$, $\sigma=0.05$, $\Lambda_\nu=20$,
$a = 0.005$, $\epsilon = 0.0005$, $N_\eta=500$, $N_x=200$, and $N_\phi=200$.
}
\end{figure}

\subsection{Initial non-Abelian noise and partial Abelianization}

In Fig.~\ref{fig4} we show results obtained from an SU(2) non-Abelian 
run in which the initial condition is taken to be a random 
superposition of discrete transverse electric field modes (both $\Pi_x$ and 
$\Pi_y$) with an ultraviolet cutoff $\Lambda_\nu = 20$
in space-time rapidity wave number $\nu$.  The amplitude 
for each mode is chosen from a Gaussian probability distribution 
centered at zero with standard deviation $\sigma = 0.03$.  As can be 
seen from Fig.~\ref{fig4}a the system very quickly generates 
chromomagnetic fields whereas during the early times energy is equally 
distributed between transverse chromoelectric and chromomagnetic fields.  
Longitudinal field energies which vanish initially
grow exponentially with a rate about twice of that of the transverse fields,
but almost saturate when the nonlinear regime is reached.
During the initial growth phase as well as in the deep nonlinear regime, 
the energy density is exponentially dominated by 
transverse chromomagnetic fields.
This again translates into the generation of 
exponentially large longitudinal pressure as shown in 
Fig.~\ref{fig4}b.  

In Fig.~\ref{fig4}c we plot various measures of the Abelianization and 
(color) correlations of the chromo-fields.  Following 
Ref.~\cite{Arnold:2004ih,Rebhan:2005re} we define a measure of the ``Abelianness''
of the field configurations through
\be
\bar C[j] = \int_0^{L_\eta} {d\eta\0L_\eta} { \left\{ \tr \left( (i[j_x,j_y])^2 \right) 
\right\}^{1/2}\0 \tr (j_x^2+j_y^2) } \, .
\ee
If the field configurations are Abelian (aligned in one color direction)
then this quantity vanishes because of the commutator in the numerator.

In order to further study the color correlations of the chromo-fields
in spatial rapidity, $\eta$, we define
\be
\chi_A(\xi)={N_c^2-1\02N_c}
\int_0^{L_\eta} {d\eta\0L_\eta} { \tr \left\{
(i[j_i(\eta+\xi),\mathcal U(\eta+\xi,\eta)j_j(\eta)])^2 \right\}
\0 \tr\{ j_k^2(\eta+\xi) \} \tr \{ j_l^2(\eta) \} } \, ,
\ee
where $\mathcal U(\eta',\eta)$ is the adjoint-representation parallel transport
from $\eta$ to $\eta'$. When colors are completely uncorrelated over a
distance $\xi$, this quantity equals unity; if they point in the
same direction, this quantity vanishes. Following Ref.~\cite{Rebhan:2005re,Arnold:2004ih} 
we define the ``Abelianization correlation length''
$\xi_A$ as the smallest distance where $\chi_A$ is larger than 1/2,
\be
\xi_A[j]=\min_{\chi_A(\xi)\ge 1/2}  (\xi) \, .
\ee

This we compare with a general correlation length, which does not focus
on color, defined through the gauge invariant function 
\be
\chi(\xi)={ \int_0^{L_\eta} {d\eta} \tr\{ j_i(\eta+\xi)
\mathcal U(\eta+\xi,\eta)j_i(\eta)\} \0
\int_0^{L_\eta} {d\eta} \tr\{ j_l(\eta) j_l(\eta)\} } \, .
\ee
This function now vanishes when fields are uncorrelated over a distance
$\xi$, and it is normalized such that $\chi(0)=1$.
We thus define the general correlation length through
\be
\xi[j]=\min_{\chi(\xi)\le 1/2}  (\xi) \, .
\ee



Fig.~\ref{fig4}c shows 
that the system becomes Abelianized with large 
color correlation length, $\xi_A[j]$, when the fields have grown such
that nonlinear self-interactions become important.
$\xi_A$ occasionally even shoots up to the size of the space-time
rapidity lattice (2.5 in this case) before settling to oscillations around 
rapidities $\sim 0.3$.
(The indication of some late-time growth of $\xi_A$ is presumably spurious,
since it is accompanied with the onset of a rapid growth of the Gauss law
violation control parameter.)
Although we show the 
output of only one run here the behavior shown is generic for all 
random seeds we have studied.

\subsection{Color-Glass-Condensate-inspired initial conditions}

In Fig.~\ref{fig5} we show results obtained by using initial 
seed fields which reflect the spectral properties
obtained by Fukushima, Gelis, and McLerran (FGM) 
within the Color-Glass-Condensate (CGC) framework
\cite{Fukushima:2006ax}. We use again a random superposition
of modes, but now involving already initially
both chromoelectric and chromomagnetic transverse fields
with a spectrum\footnote{The spectrum of fluctuation derived in
Ref.~\cite{Fukushima:2006ax} of course has also modes which are
not constant in the transverse coordinates, but in our present
framework we have to
restrict ourselves to modes which are effectively 1+1-dimensional.}
\bea
\mid\! \Pi_{\9i}(\nu)\!\mid_{\tau=\tau_0} &=& \sigma \sqrt{\nu} \nonumber \\
\mid\! A^{\9i}(\nu)\!\mid_{\tau=\tau_0} &=& \sigma /\sqrt{\nu}  \, ,
\label{FGMsketch}
\eea
for all space-time rapidity wave numbers $\nu \le \Lambda_\nu$ that
are allowed by the periodic boundary conditions of our finite $\eta$ lattice,
excluding however $\nu=0$. The phases of each color component of these
modes is taken at random, and we have used a small value $\sigma = 0.05$,
corresponding to initially weak fields.
In accordance with Ref.~\cite{Fukushima:2006ax}, the longitudinal
magnetic field is set to zero initially through $A_\eta|_{\tau=\tau_0}=0$,
but the non-Abelian Gauss law leads to nonvanishing longitudinal
chromoelectric fields even though $j^\tau\equiv 0$ initially.

With $A_\eta=0$ initially, the Gauss law constraint in the 1+1-dimensional
setting gives
\be\label{initGauss1+1}
\partial_\eta \Pi^\eta=ig[A^{\9i},\Pi_{\9i}].
\ee
Having populated the transverse field modes according to Eq.~(\ref{FGMsketch}),
we solve the lattice version of Eq.~(\ref{initGauss1+1}) to determine
the longitudinal electric field $E^{\9i}=\Pi^{\9i}/\tau$.

However, in contrast to the simpler initial conditions used above,
this presents a problem with the periodicity of our $\eta$ lattice,
since the solution thus obtained does not share the periodicity of
all other fields,
leading to a Gauss law violation at the boundary in the form of a
mismatch of $\Pi^\eta$. This initially small violation however
quickly grows and cannot be tolerated. 
We have solved this problem by singling out the lowest lying mode of $A^x$
and to calculate its contribution to the mismatch of $\Pi^\eta$.
By elementary linear algebra we determine how to rescale the color components
of this one mode such that the mismatch is eliminated, but this
rescaling is only accepted when the total amplitude of this
mode does not get modified by more than 50\%. If this is not the case,
a different set of random numbers for the phases of all transverse
color fields is generated and the procedure repeated until a configuration
is found where the amplitude of the lowest lying mode of $A_x$ is not
too far from the starting point (\ref{FGMsketch}).


\begin{figure}
\centerline
{\includegraphics[width=0.53\linewidth]{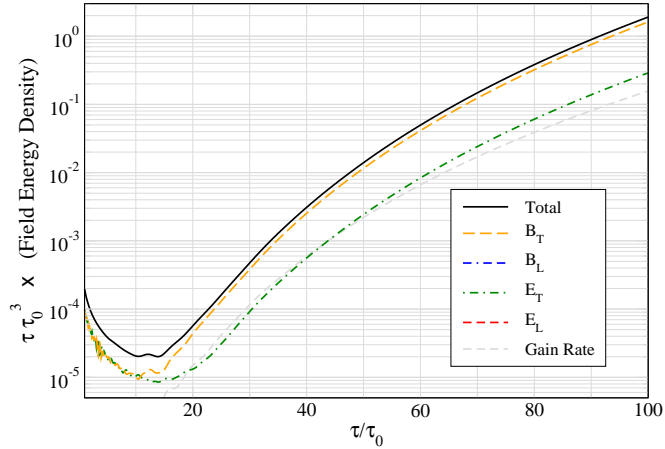}
}
\caption{\label{fig6}
Field energy density results 
from an Abelian run initialized with FGM initial conditions.  
Transverse fields and
Gauss law violation are zero to within machine precision.
The field pressure is purely longitudinal and coincides with the total
field energy density.
Run was made
using $\tau_{\rm iso}=0.1$, $\tau_0=1.0$, $m_D = 3.585$, $\sigma=0.01732$, $\Lambda_\nu=20$,
$a = 0.01$, $\epsilon = 0.001$, $N_\eta=500$, $N_x=100$, and $N_\phi=4$.
}
\end{figure}

\begin{figure}
\centerline
{\includegraphics[width=0.53\linewidth]{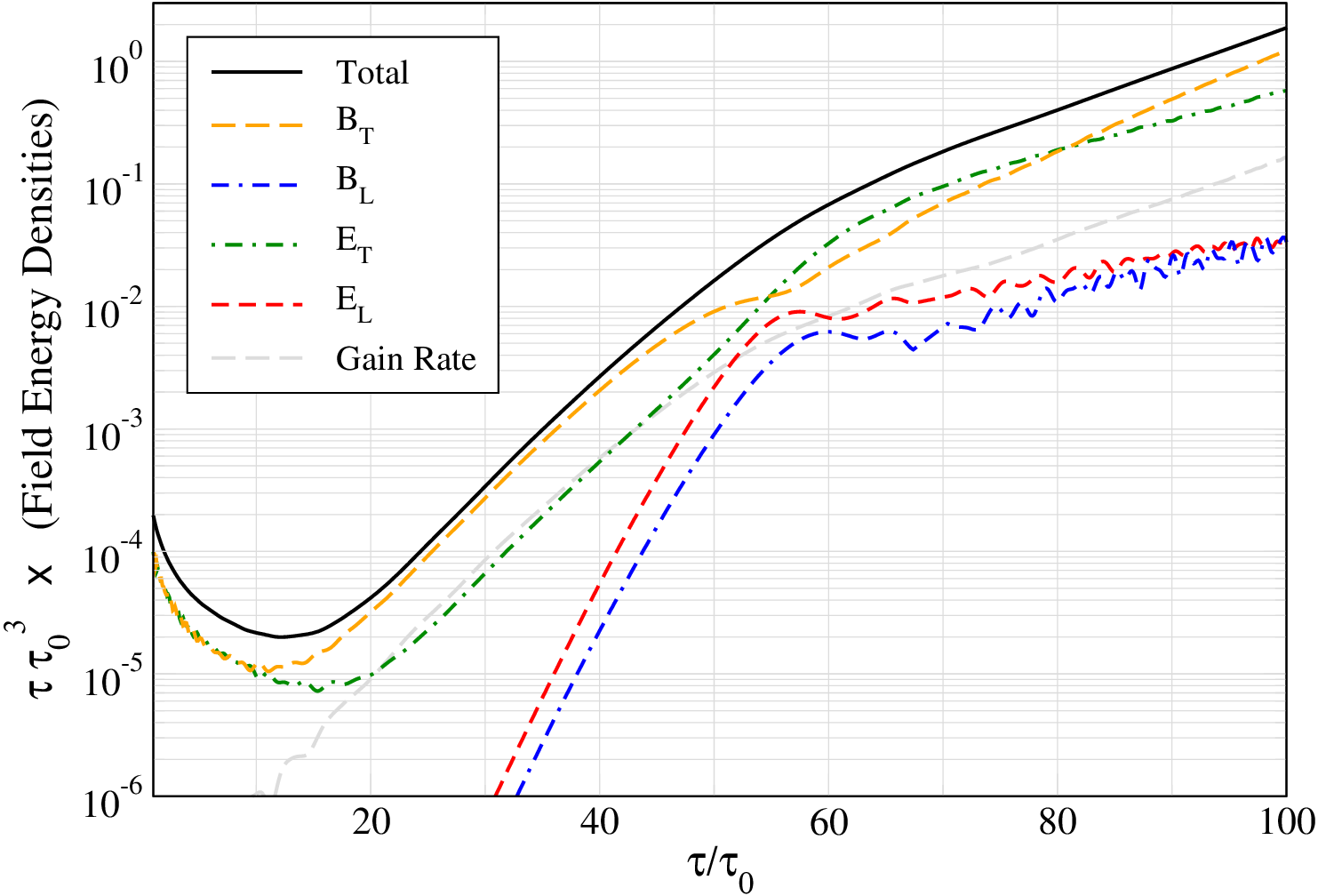}~~~~}
\vspace{1.1cm}
\centerline
{\includegraphics[width=0.5\linewidth]{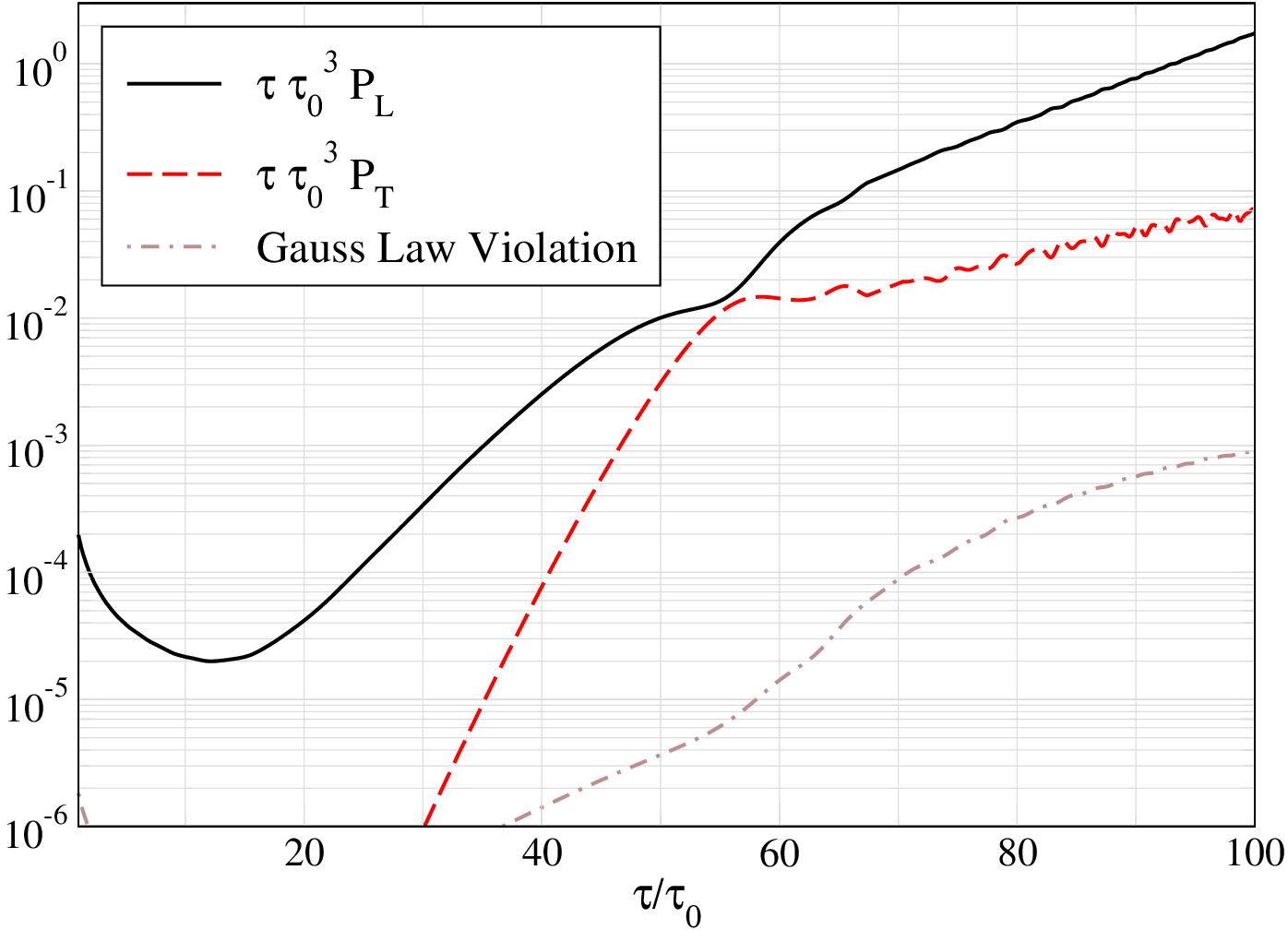}}
\vspace{1.1cm}
\centerline
{\includegraphics[width=0.5\linewidth]{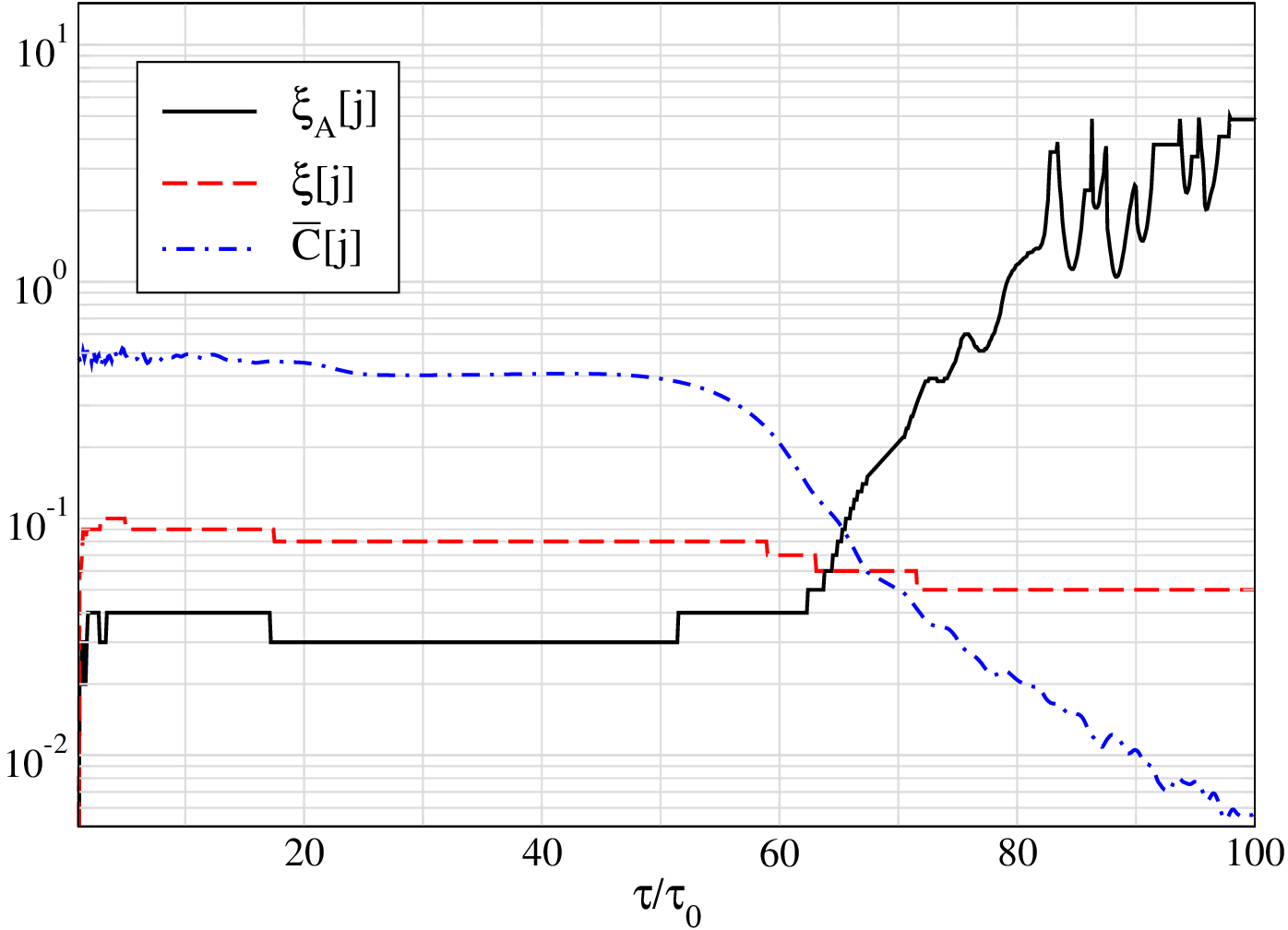}}
\caption{\label{fig7}
Results 
from non-Abelian run initialized with FGM initial conditions.  
In the top panel 
(a) we show the proper-time dependence
of the chromo-field energy densities and the energy gain rate (\ref{REG}) times an extra factor of
$\tau_0$.  In the middle panel (b) we show the 
longitudinal and transverse pressures along with our numerical 
Gauss law violation.  In bottom panel (c) we show the correlations
$\xi_A[j]$, $\xi[j]$, and ${\bar C}[j]$.  Run was made
using $\tau_{\rm iso}=0.1$, $\tau_0=1.0$, $m_D = 3.585$, $\sigma=0.01$, $\Lambda_\nu=20$,
$a = 0.01$, $\epsilon = 0.001$, $N_\eta=500$, $N_x=200$, and $N_\phi=100$.
}
\end{figure}

The results shown in Fig.~\ref{fig5} are qualitatively similar to 
those obtained with a random superposition of purely electric modes 
initial condition.  As can be seen from Fig.~\ref{fig5}a at early times 
there is equal partitioning between chromoelectric and chromomagnetic 
fields which both initially decrease and then begin to grow 
exponentially with transverse chromomagnetic fields dominating for 
nearly the entire run. At $\tau/\tau_0\sim 13$ there is a non-Abelian 
``bounce'' when the longitudinal field components become on the same 
order of magnitude as the transverse ones; however, beyond this point 
in time the transverse field components again dominate.  In 
Fig.~\ref{fig5}b we see that the field pressures which are generated 
are also similar to those obtained with a random discrete Fourier 
spectrum with the system generating an exponentially large 
longitudinal pressure due to the chromo-Weibel instability.  In 
Fig.~\ref{fig5}c the behavior of the Abelianization measure, ${\bar 
C}[j]$, and correlation lengths are again similar to the random 
discrete Fourier spectrum initial conditions showing an Abelianization 
of the fields and large color correlation length at late times.  This 
demonstrates that the qualitative features of the time evolution of 
the instability induced fields are independent of the details of the 
initial condition.\footnote{Of course, by this we mean any reasonable 
initial condition.  Choosing, for example, an initial condition which 
only had 
very high frequency modes would greatly 
delay the onset of instability driven growth of the fields.}

In Figs.~\ref{fig6} and \ref{fig7} we show results obtained using FGM 
initial conditions with a smaller Debye mass corresponding to the 
estimates of the ``gluon liberation factor'' $c$ obtained from the 
color-glass-condensate picture \cite{Kovchegov:2000hz,Lappi:2007ku}
(see Appendix \ref{appCGC} for details). 
In Fig.~\ref{fig6} we show the results obtained from an Abelian run in 
which all fields were constrained to initially point in the same 
direction in color space and in Fig.~\ref{fig7} we show the results of 
a non-Abelian SU(2) run.  As can be seen from both figures the primary 
effect of lowering $m_D$ is to slow down the growth of the 
chromo-fields; however, besides this ``stretching'' of the time axis 
there is little qualitative difference between the larger $m_D$ run 
(Fig.~\ref{fig5}) and this case, Fig.~\ref{fig7}.  We still observe 
domination by transverse chromo-fields, which now have larger color 
correlation length, and generate exponentially large longitudinal 
pressure.

\begin{figure}
\centerline{
\includegraphics[width=0.49\linewidth]{fig8-a.eps}
$\;$
\includegraphics[width=0.48\linewidth]{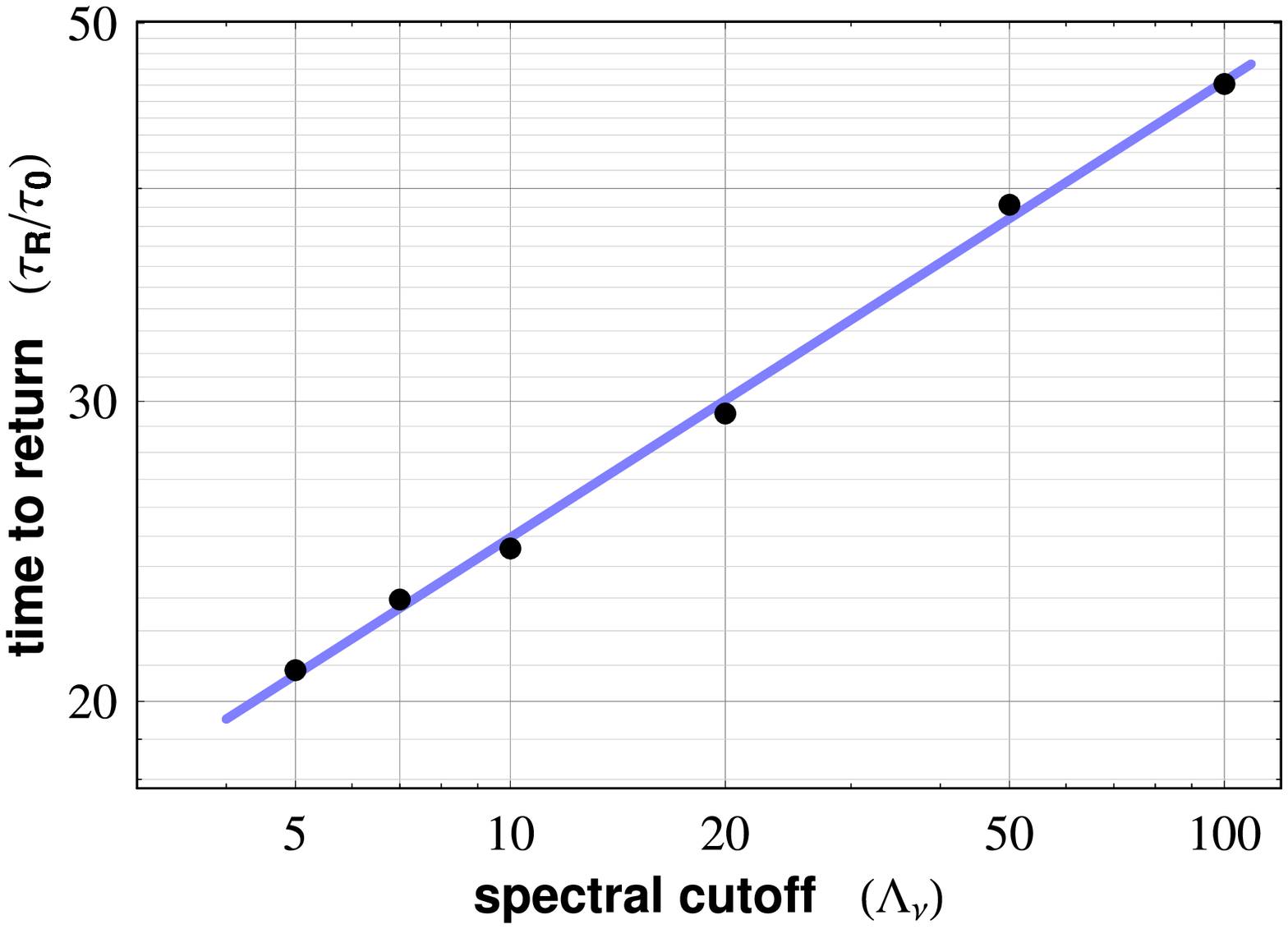}
}
\caption{\label{fig8}
Left panel (a) total field energy density results 
from a non-Abelian run initialized with FGM initial conditions and
different UV cutoffs, $\Lambda_\nu \in \{5,7,10,20,50,100\}$, imposed on the initial spectrum.  
Right panel (b) shows the ``time to return'' $\tau_R/\tau_0$, defined as the point at which
the field energy density has returned to its initial value, as a function of the FGM spectral
cutoff $\Lambda_\nu$ on a log-log plot.  Blue line shows a fit to a power law 
$\tau_R = a \tau_0 (\Lambda_\nu)^b$.
Runs were made using $\tau_{\rm iso}=0.1$, $\tau_0=1.0$, $m_D = 3.585$, $\sigma=0.06$,
$a = 0.005$, $\epsilon = 0.0025$, $N_\eta=1000$, $N_y=800$, and $N_\phi=50$.
For this figure discretization method A was used.
}
\end{figure}

In Fig.~\ref{fig8} we compare six different non-Abelian SU(2) runs 
with FGM initial conditions in which we have taken different values for 
the spectral cutoff in rapidity wave number, $\Lambda_\nu$, 
imposed on the FGM 
initial condition. As can be seen from this figure for fixed initial 
energy density the effect of increasing $\Lambda_\nu$ is to delay the 
onset of exponential growth of the chromo-fields.  This is to be 
expected since for fixed energy density the occupation number of the 
lowest $\nu$ modes must be decreased as $\Lambda_\nu$ is increased, and
higher modes have a larger delay, as already found in the Abelian
case studied in Ref.~\cite{Romatschke:2006wg}. In 
fact, the amplitude of the low-momentum modes must be decreased 
rapidly since the high-momentum modes dominate the energy density.
In Fig.~\ref{fig8}b we show a fit to the ``time to return'', $\tau_R$, of the 
scaled energy density $\tau\tau_0^3\mathcal E$, i.e.\ the time it 
takes the instability to compensate for the initial decay of the soft 
fields caused by the system's expansion. Fitting this time (in units of $\tau_0$) 
by a power-law $a (\Lambda_\nu)^b$ we find $a=13.46 \pm 0.01$ and $b=0.26 \pm 0.01$.
The exponent $b$ is consistent with being 1/4. The coefficient $a$ depends on 
the Debye mass and decreases as $m_D$ increases.

\begin{figure}
\centerline
{\includegraphics[width=0.48\linewidth]{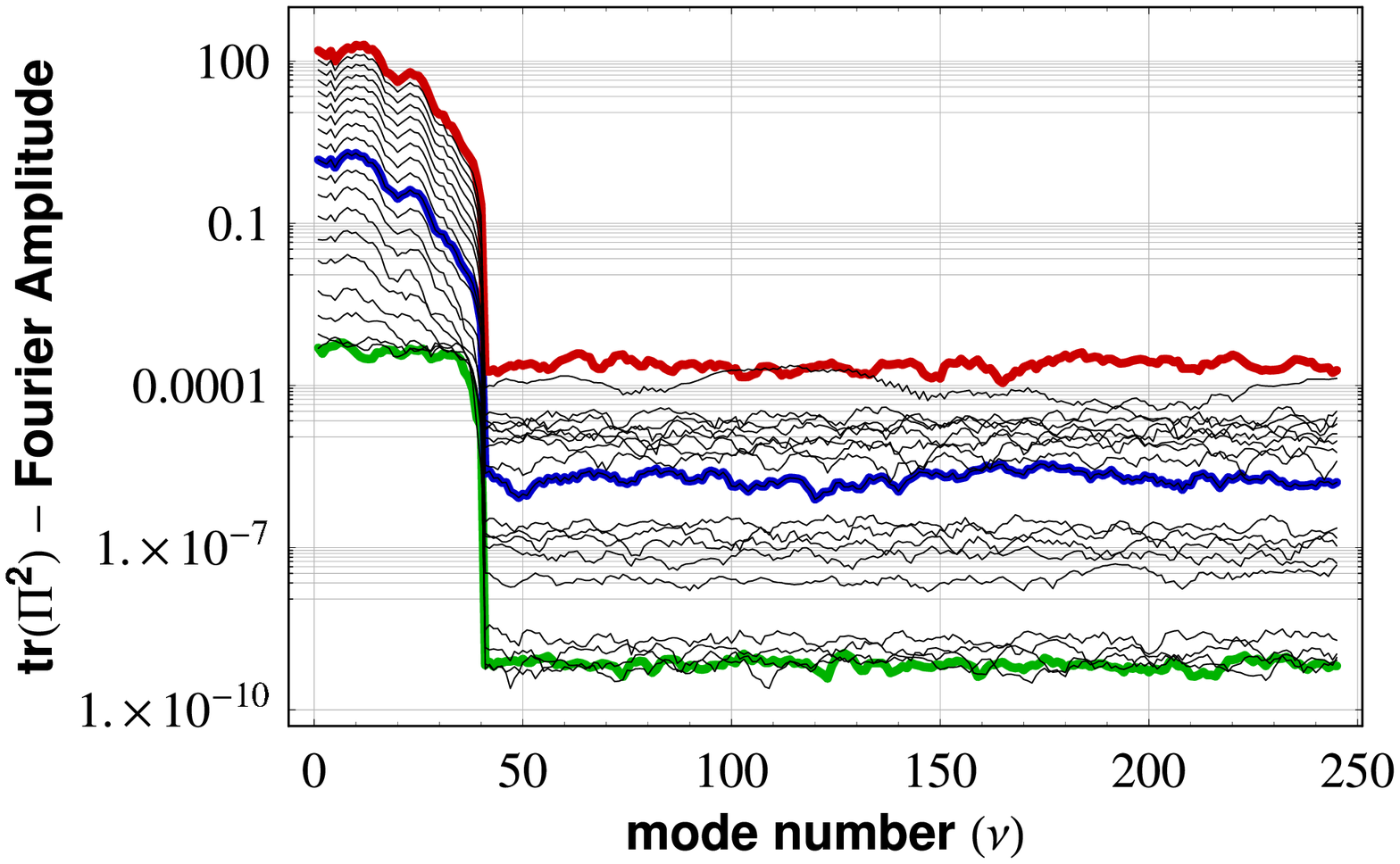}
\hspace{0.02\linewidth}
\includegraphics[width=0.48\linewidth]{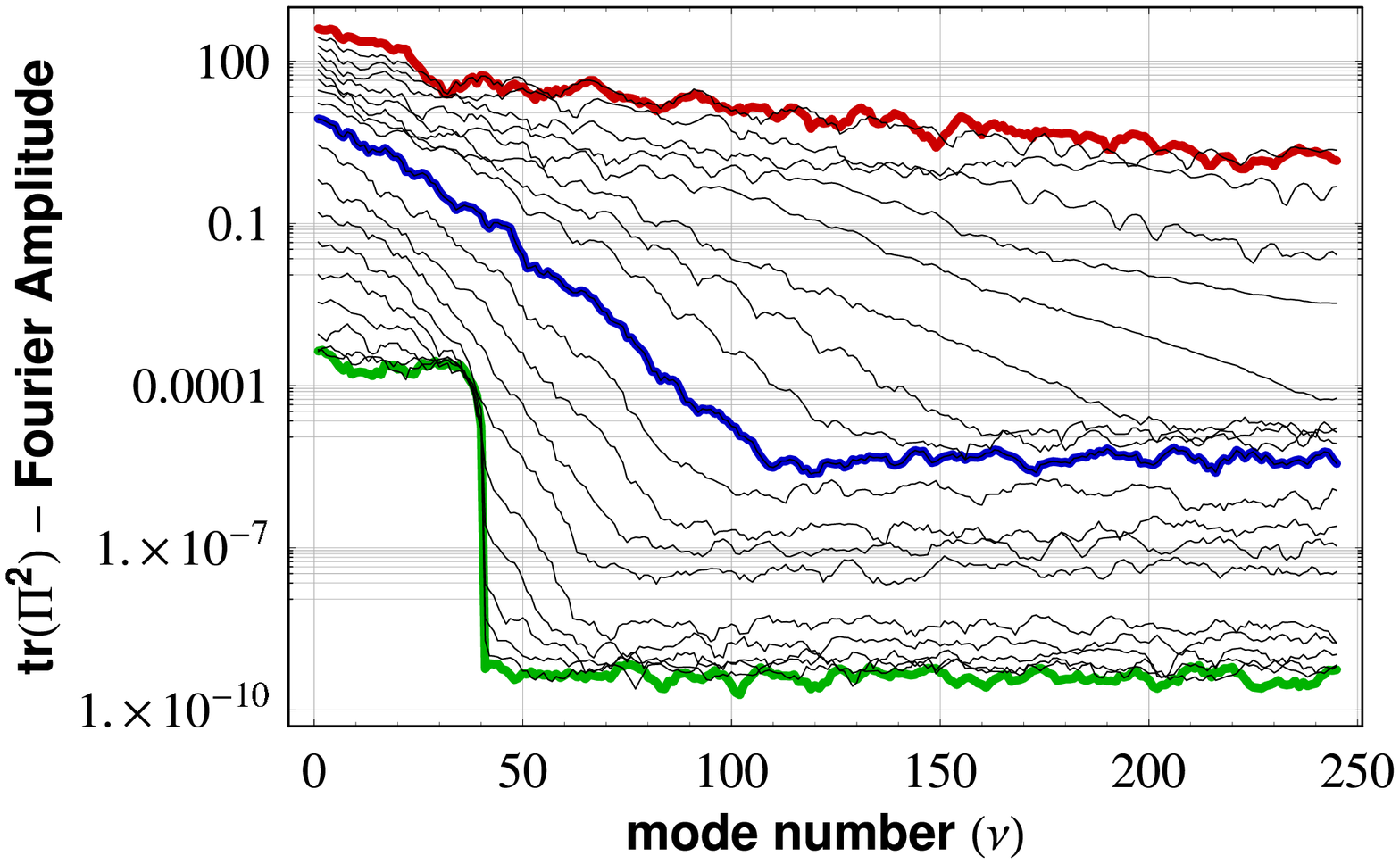}}
\caption{\label{fig9}
Fourier spectrum of the color-traced conjugate field momenta, 
${\rm tr}({\bf \Pi}^2)$, obtained
from (left) Abelian and (right) non-Abelian runs with FGM initial conditions.  
The lowest (bold green) 
line indicates the starting spectrum and the uppermost (bold red) line 
indicates the final spectrum.  In the right panel the 
bold blue line indicates the ``non-Abelian point'' at $\tau/\tau_0 \sim$  55 when
all field components become approximately the same order of magnitude.  
The Abelian and non-Abelian spectra were obtained by analyzing the currents produced during
the runs shown in Figs.~\ref{fig6} and \ref{fig7}, respectively.
}
\end{figure}

In Fig.~\ref{fig9} we compare the rapidity ($\nu$) spectrum obtained by Fourier 
transforming the trace of the conjugate field momenta, ${\rm tr}({\bf 
\Pi}^2)={\rm tr}(\Pi_{\9i}^2+\tau^2 (\Pi^\eta)^2)$, in order to 
gain more understanding of the momentum space 
dynamics of the fields in our simulations.  In the left panel, 
Fig.~\ref{fig9}a, we show the spectrum resulting from analysis of the 
induced current from the Abelian run shown in Fig.~\ref{fig6}.  In the 
right panel, Fig.~\ref{fig9}b, we show the spectrum resulting from 
analysis of the induced current from the non-Abelian run shown in 
Fig.~\ref{fig7}. The lowest (bold green) line indicates the starting spectrum, 
the bold blue line indicates the ``non-Abelian point'' at which all 
field components become approximately equal in magnitude, and the 
uppermost (bold 
red) line shows the final spectrum obtained in our simulations. As can 
be seen from this figure there is a stark qualitative difference between the 
Abelian and non-Abelian spectra with the former maintaining the 
spectral cutoff imposed on the initial condition and the latter 
``cascading'' energy to higher and higher momentum modes starting 
already at very early times. This is similar to earlier results for 
the spectra induced by instability growth \cite{Arnold:2005qs}. 
Surprisingly, in Fig.~\ref{fig9}b one sees that at the ``non-Abelian 
point'' indicated by the bold blue line that the low frequency modes 
have generated a quasi-thermal (Boltzmann) distribution up to $\nu 
\sim 80$.  In fact, the development of the quasi-thermal distribution 
begins at very early times and one can associate a temperature with 
the system by fitting the low-$\nu$ spectra with exponential fits from 
rather early times. Similar spectra are generated when one measures 
${\rm tr}({\bf A}^2)$ which also allows one to define a kind of 
magnetic temperature from that observable as well.

\section{Discussion, Conclusions, and Outlook}

In this paper we have performed the first numerical study of
non-Abelian plasma instabilities in a nonstationary, longitudinally
expanding system within the framework of discretized hard loop theory%
\footnote{Closely related instabilities have been found before numerically 
in the color-glass condensate framework in 
Ref.~\cite{Romatschke:2005pm,Romatschke:2006nk}, where the role of
plasma particles is played by high-momentum modes of the Yang-Mills field.},
extending the semi-analytical results of \cite{Romatschke:2006wg}
for the weak-field, Abelian regime.
We have worked out the case of the most unstable
modes which are constant modes in the transverse direction, making
the dynamics 1+1-dimensional in configuration space (while momentum
space remains 3-dimensional). Starting with only small rapidity fluctuations,
we found that the exponential (in $\sqrt\tau$ 
\cite{Romatschke:2006wg})
growth in the Abelian (weak-field) phase is only mildly weakened
when nonlinearities through non-Abelian self-interactions of the
collective fields set in, and this is associated with significant
degree of Abelianization in finite domains in the nonlinear regime. 
This is quite similar to what was observed
in the 1D+3V simulations in a stationary anisotropic plasma 
\cite{Rebhan:2004ur} and it remains to be seen what full
3D+3V simulations will give. However, it is quite plausible that
the 1D+3V results already capture the behavior of the more generic
3D+3V simulations, because it was recently observed \cite{Bodeker:2007fw}
that for extreme anisotropies a saturation of the
growth as was found in Refs.~\cite{Arnold:2005vb,Rebhan:2005re} at
moderate anisotropies
will occur only at correspondingly extreme values of the fields, if at all.
Indeed, our simulations start out with strong anisotropy of the
particle distribution, which rapidly grows with increasing time
according to Eq.~(\ref{xi}).

\begin{figure}[t]
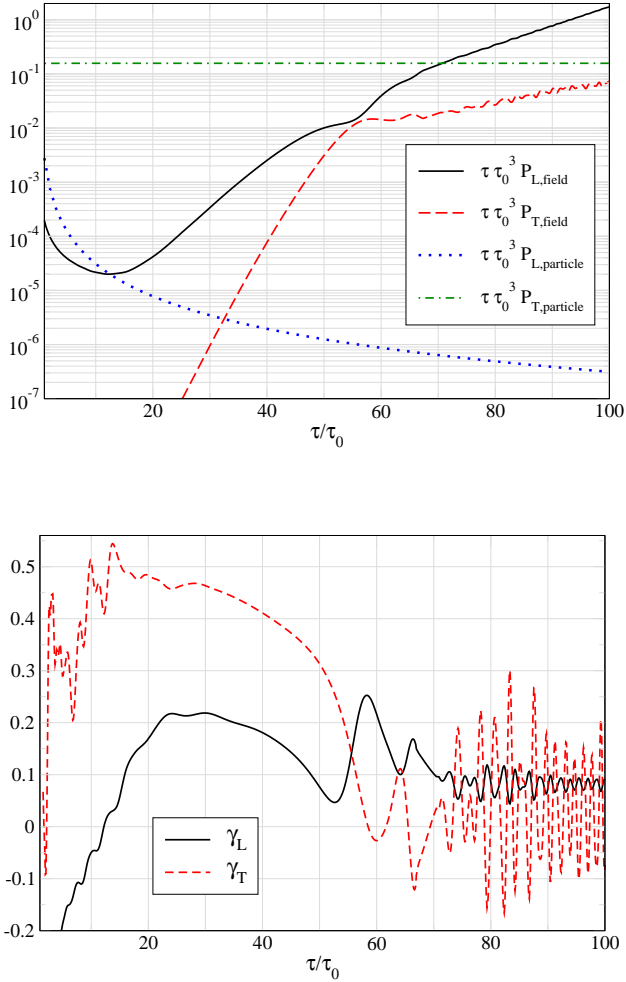

\centerline{
\includegraphics[width=0.5\linewidth]{fig10-a.eps}}
\vspace{1.1cm}
\centerline{
\includegraphics[width=0.5\linewidth]{fig10-b.eps}
}
\caption{\label{fig10}
Top panel (a) shows a comparison of particle and field pressures 
generated during a typical 
run which uses CGC-inspired ``FGM'' initial conditions.  Simulation 
parameters are the same as shown in Fig.~\ref{fig7} and the field 
pressures are the same as in panel (b) of that figure.  The particle 
pressures are obtained by evaluating the expressions given in 
Eqs.~(\ref{particlePT}) and (\ref{particlePL}).  In order to fix the 
initial energy densities we use the scheme detailed in Appendix 
\ref{appCGC} assuming $\alpha_s = 0.3$ which gives ${\cal E}(\tau_0) 
\simeq 0.3 \, Q_s^4$. 
In the lower panel (b) we show the growth rates for the field pressures
in units of $\tau_0^{-1}$.
 } 
\end{figure}

In our simulations we have found that in the non-Abelian case
the growing unstable modes tend towards a quasi-thermal spectrum
(Fig.~\ref{fig9}b) and they produce mainly longitudinal field pressure,
which grows exponentially, 
thereby realizing a bottom-up isotropization
scenario in which the soft modes make up for the strongly
decaying longitudinal particle pressure, which goes like
$1/\tau^3$. The transverse particle pressure,
which according to the CGC picture is approximately thermal
by itself, is decaying like $1/\tau$. In the hard-(expanding)-loop theory which
we have considered, we can of course only trust the beginning of
this scenario, since the backreaction of the collective fields on
the hard particle background is neglected, and it is in fact the
reservoir of energy in the hard particle background that is
feeding the growth of the soft modes, which has to stop
before the energy in the latter becomes comparable with the former.

In Fig.~\ref{fig10}a we have reproduced the results of Fig.~\ref{fig7}b
for the field pressures obtained
by choosing dimensionful parameters motivated by the CGC scenario
as described in App.~\ref{appCGC} and compared also with the
particle pressures that follow from this matching.
Notice that all quantities are multiplied by $\tau$ so that the
decaying transverse particle pressure is represented by an approximately
horizontal line. 

The time scale $\tau_0\simeq Q_s^{-1}$ can be roughly identified with
$1-1.5$ and 3 GeV for RHIC and LHC experiments, respectively, where the
plasma lifetimes are probably less than 5 fm/c $\simeq 25-35\tau_0$
for RHIC, and probably much larger than 7 fm/c $\simeq 100\tau_0$
for the LHC \cite{Eskola:2005ue}. Defining an effective growth rate of the
longitudinal pressure
by 
\be
\gamma_L=\frac{\6}{\6\tau}\ln(\tau\tau_0^3 P_{L,{\rm field}}),
\ee
we find for the example provided in Fig.~\ref{fig10} a maximal
value of about $\gamma_L\sim 0.2\,\tau_0^{-1}$ in the weak-field
regime for $\tau\gtrsim 20\,\tau_0$, and about $0.1\,\tau_0^{-1}$ in the 
strong-field (non-Abelian) regime $\tau\gtrsim 70\,\tau_0$.
This corresponds to minimum characteristic time scales of
\be
{\rm min\;}\gamma_L^{-1} \sim 
\left\{ 0.7-1\; {\rm fm/c \quad (RHIC)} \atop
0.3\; {\rm fm/c \qquad (LHC)} \right.
\ee
in the weak-field regime, and twice that in the strong-field regime.
This agrees roughly with the pre-isotropization
values obtained in Ref.~\cite{Berges:2007re}
from classical-statistical lattice gauge theory.\footnote{The higher
growth rate of the transverse field pressure, which is due to
non-Abelian self-interactions of the chromo-fields (it vanishes
in the Abelian case), is what
Ref.~\cite{Berges:2007re} would call a ``secondary'' instability.}

However, at least for the case of initially small rapidity 
fluctuations which we have considered here, there is a delay of the 
onset of plasma instabilities caused by the expansion which appears 
uncomfortably large for RHIC energies, even if one chooses smaller 
spectral cutoffs in the initial fluctuations which somewhat reduce this 
delay (cf.\ Fig.~\ref{fig8}).

Still, for the LHC our results suggest that plasma instabilities
like those studied here
will be an important phenomenon, in particular if LHC energies
make contact to a more weakly coupled quark-gluon plasma
as suggested for instance by the analysis of Ref.~\cite{Blaizot:2006tk}.
The comparison of particle and field pressure in Fig.~\ref{fig10}a
indicates upper limits for an isotropization point, which are
however strongly dependent on the initial strength of the
rapidity fluctuations. Larger seed fields will correspondingly
lower this point. However, experience from simulations of
non-Abelian plasma instabilities in the stationary anisotropic
case \cite{Arnold:2005vb,Rebhan:2005re,Bodeker:2007fw} 
lets us expect that full 3D+3V studies (or at least 2D+3V ones
\cite{Arnold:2007tr}) are required to analyse truly strong initial fields.
This will be the subject of follow-up work.

\section*{Acknowledgements} 
We thank Peter Arnold, J\"urgen Berges, Paul Romatschke, 
Kari Rummukainen, and Christof Wetterich for useful conversations and 
feedback.  M.S.\ was supported by DFG project GR 1536/6-1 and the 
Kavli Institute for Theoretical Physics NSF Grant No. PHY05-51164; 
A.R.\ and M.A.\ acknowledge support by FWF project P19526.  M.S.\ 
and M.A.\ also acknowledge support during the Galileo Galilei Institute for 
Theoretical Physics program on High Density QCD.

\newcommand{\inv}[1]{\frac{1}{#1}}

\appendix

\section{Lattice Discretization Method A}
\label{methodAapp}

The one-dimensional situation studied herein assumes that fields vary 
only in the $\eta$-direction. We then have transverse adjoint scalar 
fields $A^i \, \textrm{with}\, i=1,2$ and auxiliary fields 
${\cal W}_{\phi,\bar y}$ 
which all are defined on the sites $s$ of a periodic spatial rapidity
lattice with (dimensionless) lattice 
spacing $a$. The conjugate momenta $\Pi_i$ live on the 
temporal links, while the conjugate momentum $\Pi^\eta$
of the gauge field $A_\eta$,
which appears only in the form of a parallel transporter
$U_{s + \inv 2} = \exp \left(iga A_{\eta,s}\right)$,
will be treated as located on the timelike plaquette on top
of the link between site $s$ and $s+1$.

Apart from $U$, all of these fields are represented by $N_c \times 
N_c$ traceless Hermitian matrices which for SU(2) reduce to the 
$2\times 2$ Pauli matrices. Although we are going to make explicit
all occurrences of the coupling $g$, in practice we have taken $g=1$ through a 
rescaling of the fields. 

Covariant derivatives are defined in three versions: left- and right-covariant,
\begin{align}
D^R_\eta A^\alpha_s \equiv
\frac{A^\alpha_s - U_{s - \inv 2} A^\alpha_{s-1} U_{s - \inv 2}^\dagger}{a}
\, , \quad \quad
D^L_\eta A^\alpha_s \equiv
\frac{U_{s + \inv 2}^\dagger A^\alpha_{s+1} U_{s + \inv 2} - A^\alpha_s}{a} \,,
\end{align}
and symmetric,
\begin{align}
 D^S_\eta \equiv (D^L_\eta + D^R_\eta)/{2} \,.
\end{align}
The second-order is given by
\begin{align}
D_\eta^2 \equiv ({D^L_\eta - D^R_\eta})/{a} \,
\end{align}
and is automatically symmetric.

In method A,
the auxiliary field $\mathcal W(\tau,\eta;\phi,y)$ of the continuum theory
is modelled by a large number of fields $\mathcal W_{s;\phi,\bar y}$
with $\bar y=y-\eta$ discretized with $N_{\bar y}$ points in the
interval $(-\Lambda_{\bar y},\Lambda_{\bar y})$ and $N_\phi$ points
for $0\le\phi<2\pi$. Additionally, we can absorb all or part of the
denominator appearing in Eq.~(\ref{current}) for the induced current
by writing
\be
\bar {\mathcal W}_{s;\phi,\bar y}(\tau)=f^{-1}(\tau,\bar y){\mathcal W}_{s;\phi,\bar y}(\tau)
\ee
with
\begin{align}
f (\tau, \bar y) = \left( 1 + \frac{\tau^2}{\tau_{\mathrm{iso}}^2} 
\sinh^2(\bar y) \right)^{\lambda} \,,
\end{align}
with $\lambda$ a number between 0 and 2. This does not produce
extra terms in the equation of motion for $\mathcal W$ because
\be
\left[ \6_\tau + \frac{\tanh(\bar y)}{\tau}\6_\eta \right] f = 0.
\ee

The
induced current (\ref{current}) is obtained from the auxiliary fields (which
at $\tau=\tau_0=1$ are taken to vanish) according to
\be
j_s^\alpha(\eta) = - \frac{m^2_D \Lambda_{\bar y}}{N_\phi N_{\bar y}}
\sum_{\phi}\sum_{\bar y}
{V^\alpha} f^{\lambda-2} \bar{\cal W}_{s;\phi,\bar y}(\tau) \,,
\ee
with $V^\alpha$ defined in (\ref{velocityDef1}).

The equations of motion of the various fields are then solved numerically
by a leapfrog procedure.
The first step is to calculate the conjugate momenta from
\begin{align}\label{app_equ_motion}
 \Pi_{i,s} (\tau + \frac\epsilon2 ) &= 
 \Pi_{i,s} (\tau - \frac\epsilon2 ) + \epsilon \left(
\tau j^i_s 
+ \inv{\tau} D_\eta^2  A^i_s
+ \tau g^2 i[A^{j}_s, i[A^{j}_s, A^i_s]] \right)_\tau 
\,, \nonumber \\
\Pi^\eta_s (\tau + \frac\epsilon2) &= 
\Pi^\eta_s (\tau - \frac\epsilon2 ) + \epsilon \left( - \frac{\tau}{2}
(j^\eta_s + U_{s + \inv 2}^\dagger j^\eta_{s + 1} U_{s + \inv 2} )
+ \frac{ ig}{\tau} [A^i_s, D_\eta^L A^i_s]  \right)_\tau \,.
\end{align}
The second step is to update the fields according to
\bea
A^i_s(\tau + \epsilon) &=& A^i_s(\tau) + \epsilon 
(\tau + \frac \epsilon 2)^{-1}
\Pi_{i,s}(\tau + \frac \epsilon 2) \,,\\
U_{s + \inv 2} (\tau + \epsilon) &=&
\exp \left(ig \epsilon a (\tau + \frac \epsilon 2) \Pi^\eta_s (\tau + \frac \epsilon 2
)\right)
U_{s + \inv 2} (\tau) \,,
\eea
and the auxiliary fields $\bar{\cal W}$ according to
\begin{align}
\bar {\cal  W}_{s;\phi,\bar y}(\tau + \epsilon) &=
\bar {\cal W}_{s;\phi,\bar y}(\tau - \epsilon)
+ 2 \epsilon 
\Bigl\{f(\tau,\bar y)^{-1}{\cal C}
- g \frac{{\cal A}}{\cosh(\bar y)} \nonumber \\
&\quad + \frac{\tanh(\bar y)}{\tau}
\left[ \left(1-\frac{\tau^2}{\tau_{\mathrm iso}^2}\right) f(\tau,\bar y)^{-1}
{\cal B}
- (D_\eta^S - \partial_{\bar y}) \bar {\cal W}_{s;\phi,\bar y}(\tau)
\right] \Bigr\}
\end{align}
with
\begin{align}
{\cal A} &\equiv i [v^i A^i_s(\tau), \bar {\cal W}_{s;\phi,\bar y}(\tau)]
\,, \qquad
{\cal B} \equiv v^i D_\eta^S A^i_s(\tau) \,, \nonumber \\ 
{\cal C} &\equiv \left[ (\tau + \frac \epsilon 2)^{-1} {v^i}\Pi_{i,s} (\tau + \frac{\epsilon}{2}) + (\tau - \frac \epsilon 2)^{-1} {v^i}\Pi_{i,s} (\tau - \frac{\epsilon}{2}) \right]\big/2
\nonumber \\ &\quad
- \frac{ \sinh(\bar y)}{4 \tau^2_{\mathrm{iso}}} \Bigr\{
(\tau + \frac \epsilon 2)^2 
\left[\Pi^\eta_s(\tau + \frac{\epsilon}{2}) +
       U_{s-\frac{1}{2}} \Pi^\eta_{s-1}(\tau + \frac{\epsilon}{2}) U_{s-\frac{1}{2}}^\dagger\right] \nonumber \\ & \qquad\qquad\quad
+(\tau - \frac \epsilon 2)^2 
\left[\Pi^\eta_s(\tau - \frac{\epsilon}{2}) +
       U_{s-\frac{1}{2}} \Pi^\eta_{s-1}(\tau - \frac{\epsilon}{2}) U_{s-\frac{1}{2}}^\dagger\right]\Bigr\}\,.
\end{align}

The Gauss law constraint is checked by evaluating
\be
\frac1{N_\eta}\sum_s \tr\left(
\inv \tau D_\eta^S \Pi^\eta_s (\tau + \frac{\epsilon}{2})
- \frac{ig}{\tau} [A^i_s(\tau), \Pi_{i,s}(\tau + \frac{\epsilon}{2})]
-j^\tau(\tau)
\right)^2 \, ,
\label{gausslawlattice}
\ee
and then taking a square root to obtain the results shown reported
in the main body of the text.

\section{Lattice Discretization Method B}
\label{methodBapp}

In method B, the shifted momentum space rapidity $\bar y = y - \eta$
is not discretized with uniform spacing in $\bar y$ but
in a velocity-like variable $x$, $-1 < x < 1$, defined by
\begin{align}
\bar y \equiv {\rm atanh }(x)\,,\quad d\bar y = \inv{1 - x^2} dx.
\end{align}
Compared to method A, this has the effect of giving more lattice
points around $\bar y = 0$, where the $\bar {\cal W}$ functions 
are typically sharply peaked.

With $\sinh^2(\bar y) = {x^2}/({1 - x^2})$ and
$\cosh^2(\bar y) = 1/({1 - x^2})$
this leads to
\begin{align}
\bar {\cal  W}_{s;\phi,x}(\tau + \epsilon) &=
\bar {\cal W}_{s;\phi,x}(\tau - \epsilon)
+ 2 \epsilon \Bigg( f(\tau,x)^{-1}\,{\cal C} - g (1 - x^2)^{\frac{1}{2}} {\cal A} \nonumber\\
&\quad + \frac{x}{\tau} \left[
\left(1 - \frac{\tau^2}{\tau^2_{\textrm{iso}}} \right) f(\tau,x)^{-1}\,{\cal B}
- ( D_\eta^S - (1 - x^2) \partial_x) \bar {\cal W}_{s;\phi,x}(\tau)
\right]\Bigg)
\end{align}
with (choosing now $\lambda=2$)
\be
f(\tau,x) \equiv \left(1 -x^2\right)^{-2}
\left(1 + (\frac{\tau^2}{\tau^2_{\textrm{iso}}} -
1)x^2\right)^{2} \, , 
\ee
and
\begin{align}
{\cal A} &\equiv i [v^i A^i_s(\tau), \bar {\cal W}_{s;\phi,x}(\tau) ]
\,, \qquad
{\cal B} \equiv v^i D^S_\eta A^i_s(\tau) \,, \nonumber \\
{\cal C} &\equiv \frac{v^i}{2\tau} \left[ 
\Pi_{i,s} (\tau - \frac{\epsilon}{2}) +
\Pi_{i,s} (\tau + \frac{\epsilon}{2})
\right] \nonumber \\
& \hspace{1cm} - \frac{\tau^2x}{4 \tau^2_{\textrm{iso}} (1 - x^2)^{\frac{1}{2}}}
\left[ \Pi^\eta_s(\tau - \frac{\epsilon}{2}) +
       U_{s-\frac{1}{2}} \Pi^\eta_{s-1}(\tau - \frac{\epsilon}{2}) U_{s-\frac{1}{2}}^\dagger
       \right. \nonumber \\
       & \left. \hspace{3cm}
       + \Pi^\eta_s(\tau + \frac{\epsilon}{2}) +
       U_{s-\frac{1}{2}} \Pi^\eta_{s-1}(\tau + \frac{\epsilon}{2}) U_{s-\frac{1}{2}}^\dagger
\right] \,.
\end{align}
The currents are then given by
\begin{align}
j^\tau &= -m_D^2 \int_0^{2 \pi} d\phi \int_{-1}^1 dx \, (1 - x^2)^{-\frac{3}{2}}
 \, \bar {\cal W} \,, \nonumber \\
j^i &= -m_D^2 \int_0^{2 \pi} d\phi \int_{-1}^1 dx \, v^i \, (1 - x^2)^{-1}
 \, \bar {\cal W} \,, \nonumber \\
j^\eta &= -m_D^2 \int_0^{2 \pi} d\phi \int_{-1}^1 dx \, x \, (1 - x^2)^{-\frac{3}{2}}
 \, \bar {\cal W} \,,
\end{align}
where the integrations over $x$ and $\phi$ are replaced by uniformly spaced discrete sums.

All other lattice equations of motion are as in method A.

\section{Matching to CGC parameters}\label{appCGC}

For fixing the dimensionful parameters of our numerical
simulation in a way
that makes contact with heavy-ion physics, 
we proceed as in Ref.~\cite{Romatschke:2006wg}
and refer to the Color-Glass-Condensate framework
\cite{McLerran:1993ni,Iancu:2003xm} and take as starting
time for the plasma phase $\tau_0\simeq
Q_s^{-1}$, where $Q_s$ is the so-called saturation scale.

In order to determine the only other dimensionful parameter
in our HEL effective field equations, the Debye mass
$m_D$ at the (fictitious because pre-plasma) time $t_{\rm iso}$,
we assume a squashed Bose-Einstein distribution function for the
hard particle distribution function (\ref{faniso}) through
$f_{\rm iso}(p)=\mathcal N (2N_g) / (e^{p/T}-1)$ where $N_g=N_c^2-1$ 
is the number
of gluons and $\mathcal N$ a normalization
that is adjusted such that at $\tau=\tau_0$ the hard-gluon density
of CGC estimates is matched. 
Since the expansion is by assumption purely longitudinal, $T$ is a constant
transverse temperature, and it has indeed been found in
CGC calculations that the gluon distribution is approximately thermal
in the transverse directions, with $T=Q_s/d$ and $d^{-1}\simeq 0.47$
according to Ref.~\cite{Iancu:2003xm}.
The normalization $\mathcal N$ can then be fixed by
following Ref.~\cite{Baier:2002bt}, who
write the initial hard-gluon density as
\be
n(\tau_0)=c \frac{N_g Q_s^3}{4\pi^2 N_c\alpha_s (Q_s \tau_0)},
\ee 
where $c$ is
the gluon liberation factor, for which different estimates can be extracted
from the literature. 

According to Ref.~\cite{Baier:2002bt},
the numerical CGC simulations of Ref.~\cite{Krasnitz:2001ph,Krasnitz:2003jw}
correspond to $c\simeq 0.5$, while an approximate analytical calculation
by Kovchegov \cite{Kovchegov:2000hz} gave $c=2\ln2\approx 1.386$. 
We adopted this higher
value for the numerical simulations in Figs.~\ref{fig6}-\ref{fig9},
which is the more optimistic one from the point of view of
plasma instabilities and which is actually not far from
the most recent numerical result $c\simeq 1.1$
by Lappi \cite{Lappi:2007ku}. 

With $\tau_{\rm iso}$ remaining a free parameter which determines how
anisotropic the gluon distribution is at $\tau_0$, 
the normalization $\mathcal N$ is now fixed by 
\be
n(\tau_0)\frac{\tau_0}{\tau_{\rm iso}}=
n(\tau_{\rm iso})=\frac{2 \zeta(3)}{\pi^2}\mathcal N N_g T^3.
\ee
For a purely gluonic plasma, the isotropic Debye mass is given by
\be
m_D^2(\tau_{\rm iso})=\mathcal N \frac{4\pi \alpha_s N_c T^2}{3},
\ee
which together leads to
\be
m_D^2(\tau_{\rm iso})\tau_0^2(Q_s \tau_0)^{-1}= \frac{\pi c d }{6 \zeta(3)}
\frac{\tau_0}{\tau_{\rm iso}} 
\approx 1.285
\frac{\tau_0}{\tau_{\rm iso}},
\ee
when $c=2\ln 2$ and $N_c=3$. We adopt this value for our simulations
where $N_c=2$, since in previous studies of the stationary anisotropic
situation little difference was found between the SU(2) and the SU(3) case
provided $m_D$ was the same \cite{Rebhan:2005re}.
With our choice of an initial anisotropy given by $\tau_0/\tau_{\rm iso}=10$,
equating $\tau_0=Q_s^{-1}$ and using units where $\tau_0=1$,
the above result corresponds to the value $m_D=3.585$ employed in
Figs.~\ref{fig6}-\ref{fig9}.

The lower value $c\simeq 0.5$ for the gluon liberation factor
corresponds to a smaller Debye mass,
which turns out to be
rather expensive in computer time, because one has then to
go to much larger values of $\tau$ to obtain comparable effects and one
cannot increase the time steps much without losing accuracy.
However, in order to see the effect of this lower value of $c$
(which now seems disfavored \cite{Lappi:2007ku}), it should
suffice to simply rescale the $\tau$ values of Figs.~\ref{fig6}-\ref{fig9}
such that the weak-field Abelian regime matches the semi-analytical
results presented in Fig.~1 of Ref.~\cite{Romatschke:2006wg},
where $c=0.5$ was employed.



\begin{thebibliography}{10}
\newcommand{\enquote}[1]{``#1''}

\bibitem{Tannenbaum:2006ch}
M.~J. Tannenbaum, \enquote{{Recent results in relativistic heavy ion
  collisions: From 'a new state of matter' to 'the perfect fluid'}}, {\it Rept.
  Prog. Phys.\/} {\bf 69}, 2005 (2006).

\bibitem{Teaney:2003kp}
D.~Teaney, \enquote{{Effect of shear viscosity on spectra, elliptic flow, and
  {H}anbury {B}rown-{T}wiss radii}}, {\it Phys. Rev.\/} {\bf C68}, 034913
  (2003).

\bibitem{Romatschke:2007mq}
P.~Romatschke and U.~Romatschke, \enquote{{Viscosity Information from
  Relativistic Nuclear Collisions: How Perfect is the Fluid Observed at
  {RHIC}?}}, {\it Phys. Rev. Lett.\/} {\bf 99}, 172301 (2007).

\bibitem{Song:2007ux}
H.~Song and U.~W. Heinz, \enquote{{Causal viscous hydrodynamics in 2+1
  dimensions for relativistic heavy-ion collisions}}, arXiv:0712.3715 [nucl-th]
  (2007).

\bibitem{Kovtun:2004de}
P.~Kovtun, D.~T. Son and A.~O. Starinets, \enquote{Viscosity in strongly
  interacting quantum field theories from black hole physics}, {\it Phys. Rev.
  Lett.\/} {\bf 94}, 111601 (2005).

\bibitem{Wong:1996va}
S.~M.~H. Wong, \enquote{Thermal and chemical equilibration in relativistic
  heavy ion collisions}, {\it Phys. Rev.\/} {\bf C54}, 2588 (1996).

\bibitem{Wong:1997dv}
S.~M.~H. Wong, \enquote{{$\alpha_s$} dependence in the equilibration in
  relativistic heavy ion collisions}, {\it Phys. Rev.\/} {\bf C56}, 1075
  (1997).

\bibitem{Baier:2000sb}
R.~Baier, A.~H. Mueller, D.~Schiff and D.~T. Son, \enquote{`{B}ottom-up'
  thermalization in heavy ion collisions}, {\it Phys. Lett.\/} {\bf B502}, 51
  (2001).

\bibitem{Arnold:2003rq}
P.~Arnold, J.~Lenaghan and G.~D. Moore, \enquote{{QCD} plasma instabilities and
  bottom-up thermalization}, {\it JHEP\/} {\bf 08}, 002 (2003).

\bibitem{Mrowczynski:1988dz}
S.~Mr{\'o}wczy{\'n}ski, \enquote{Stream instabilities of the quark-gluon
  plasma}, {\it Phys. Lett.\/} {\bf B214}, 587 (1988).

\bibitem{Pokrovsky:1988bm}
Y.~E. Pokrovsky and A.~V. Selikhov, \enquote{Filamentation in a quark-gluon
  plasma}, {\it JETP Lett.\/} {\bf 47}, 12 (1988).

\bibitem{Mrowczynski:1993qm}
S.~Mr{\'o}wczy{\'n}ski, \enquote{Plasma instability at the initial stage of
  ultrarelativistic heavy ion collisions}, {\it Phys. Lett.\/} {\bf B314}, 118
  (1993).

\bibitem{Mueller:2005un}
A.~H. Mueller, A.~I. Shoshi and S.~M.~H. Wong, \enquote{A possible modified
  'bottom-up' thermalization in heavy ion collisions}, {\it Phys. Lett.\/} {\bf
  B632}, 257 (2006).

\bibitem{Bodeker:2005nv}
D.~B{\"o}deker, \enquote{The impact of {QCD} plasma instabilities on bottom-up
  thermalization}, {\it JHEP\/} {\bf 10}, 092 (2005).

\bibitem{Arnold:2005qs}
P.~Arnold and G.~D. Moore, \enquote{The turbulent spectrum created by
  non-abelian plasma instabilities}, {\it Phys. Rev.\/} {\bf D73}, 025013
  (2006).

\bibitem{Mueller:2006up}
A.~H. Mueller, A.~I. Shoshi and S.~M.~H. Wong, \enquote{On {K}olmogorov wave
  turbulence in {QCD}}, {\it Nucl. Phys.\/} {\bf B760}, 145 (2007).

\bibitem{Arnold:2007cg}
P.~Arnold and G.~D. Moore, \enquote{{Non-Abelian Plasma Instabilities for
  Extreme Anisotropy}}, {\it Phys. Rev.\/} {\bf D76}, 045009 (2007).

\bibitem{Asakawa:2006tc}
M.~Asakawa, S.~A. Bass and B.~M{\"u}ller, \enquote{Anomalous viscosity of an
  expanding quark-gluon plasma}, {\it Phys. Rev. Lett.\/} {\bf 96}, 252301
  (2006).

\bibitem{Weibel:1959}
E.~S. Weibel, \enquote{Spontaneously growing transverse waves in a plasma due
  to an anisotropic velocity distribution}, {\it Phys. Rev. Lett.\/} {\bf 2},
  83 (1959).


\bibitem{Romatschke:2003ms}
P.~Romatschke and M.~Strickland, \enquote{Collective modes of an anisotropic
  quark gluon plasma}, {\it Phys. Rev.\/} {\bf D68}, 036004 (2003).

\bibitem{Romatschke:2004jh}
P.~Romatschke and M.~Strickland, \enquote{Collective modes of an anisotropic
  quark-gluon plasma. {II}}, {\it Phys. Rev.\/} {\bf D70}, 116006 (2004).

\bibitem{Schenke:2006xu}
B.~Schenke, M.~Strickland, C.~Greiner and M.~H. Thoma, \enquote{{A model of the
  effect of collisions on QCD plasma instabilities}}, {\it Phys. Rev.\/} {\bf
  D73}, 125004 (2006).

\bibitem{Schenke:2006fz}
B.~Schenke and M.~Strickland, \enquote{{Fermionic collective modes of an
  anisotropic quark-gluon plasma}}, {\it Phys. Rev.\/} {\bf D74}, 065004
  (2006).

\bibitem{Rebhan:2004ur}
A.~Rebhan, P.~Romatschke and M.~Strickland, \enquote{Hard-loop dynamics of
  non-abelian plasma instabilities}, {\it Phys. Rev. Lett.\/} {\bf 94}, 102303
  (2005).

\bibitem{Blaizot:2001nr}
J.-P. Blaizot and E.~Iancu, \enquote{The quark-gluon plasma: Collective
  dynamics and hard thermal loops}, {\it Phys. Rept.\/} {\bf 359}, 355 (2002).

\bibitem{Pisarski:1997cp}
R.~D. Pisarski, \enquote{Nonabelian {D}ebye screening, tsunami waves, and
  worldline fermions}, hep-ph/9710370 (1997).

\bibitem{Mrowczynski:2000ed}
S.~Mr{\'o}wczy{\'n}ski and M.~H. Thoma, \enquote{Hard loop approach to
  anisotropic systems}, {\it Phys. Rev.\/} {\bf D62}, 036011 (2000).

\bibitem{Mrowczynski:2004kv}
S.~{Mr\'owczy\'nski}, A.~Rebhan and M.~Strickland, \enquote{Hard-loop effective
  action for anisotropic plasmas}, {\it Phys. Rev.\/} {\bf D70}, 025004 (2004).

\bibitem{Arnold:2004ih}
P.~Arnold and J.~Lenaghan, \enquote{The abelianization of {QCD} plasma
  instabilities}, {\it Phys. Rev.\/} {\bf D70}, 114007 (2004).

\bibitem{Arnold:2005vb}
P.~Arnold, G.~D. Moore and L.~G. Yaffe, \enquote{The fate of non-abelian plasma
  instabilities in 3+1 dimensions}, {\it Phys. Rev.\/} {\bf D72}, 054003
  (2005).

\bibitem{Rebhan:2005re}
A.~Rebhan, P.~Romatschke and M.~Strickland, \enquote{Dynamics of quark-gluon
  plasma instabilities in discretized hard-loop approximation}, {\it JHEP\/}
  {\bf 0509}, 041 (2005).

\bibitem{Arnold:2005ef}
P.~Arnold and G.~D. Moore, \enquote{{QCD} plasma instabilities: The nonabelian
  cascade}, {\it Phys. Rev.\/} {\bf D73}, 025006 (2006).

\bibitem{Bodeker:2007fw}
D.~B{\"o}deker and K.~Rummukainen, \enquote{{Non-abelian plasma instabilities
  for strong anisotropy}}, {\it JHEP\/} {\bf 07}, 022 (2007).

\bibitem{Romatschke:2006wg}
P.~Romatschke and A.~Rebhan, \enquote{Plasma instabilities in an
  anisotropically expanding geometry}, {\it Phys. Rev. Lett.\/} {\bf 97},
  252301 (2006).

\bibitem{Romatschke:2005pm}
P.~Romatschke and R.~Venugopalan, \enquote{Collective non-abelian instabilities
  in a melting color glass condensate}, {\it Phys. Rev. Lett.\/} {\bf 96},
  062302 (2006).

\bibitem{Romatschke:2006nk}
P.~Romatschke and R.~Venugopalan, \enquote{The unstable glasma}, {\it Phys.
  Rev.\/} {\bf D74}, 045011 (2006).

\bibitem{Iancu:2003xm}
E.~Iancu and R.~Venugopalan, \enquote{The color glass condensate and high
  energy scattering in {QCD}}, in \enquote{Quark-gluon plasma 3}, eds. R.~C.
  Hwa and X.-N. Wang, pp. 249--336 (World Sci., Singapore, 2003).

\bibitem{Arnold:2007tr}
P.~Arnold and P.-S. Leang, \enquote{{Lessons from non-Abelian plasma
  instabilities in two spatial dimensions}}, {\it Phys. Rev.\/} {\bf D76},
  065012 (2007).

\bibitem{Dumitru:2005gp}
A.~Dumitru and Y.~Nara, \enquote{{QCD} plasma instabilities and
  isotropization}, {\it Phys. Lett.\/} {\bf B621}, 89 (2005).

\bibitem{Dumitru:2006pz}
A.~Dumitru, Y.~Nara and M.~Strickland, \enquote{Ultraviolet avalanche in
  anisotropic non-abelian plasmas}, {\it Phys. Rev.\/} {\bf D75}, 025016
  (2007).

\bibitem{Dumitru:2007rp}
A.~Dumitru, Y.~Nara, B.~Schenke and M.~Strickland, \enquote{{Jet broadening in
  unstable non-Abelian plasmas}}, arXiv:0710.1223 [hep-ph] (2007).

\bibitem{Berges:2007re}
J.~Berges, S.~Scheffler and D.~Sexty, \enquote{{Bottom-up isotropization in
  classical-statistical lattice gauge theory}}, arXiv:0712.3514 [hep-ph]
  (2007).

\bibitem{Taylor:1990ia}
J.~C. Taylor and S.~M.~H. Wong, \enquote{The effective action of hard thermal
  loops in {QCD}}, {\it Nucl. Phys.\/} {\bf B346}, 115 (1990).

\bibitem{Braaten:1992gm}
E.~Braaten and R.~D. Pisarski, \enquote{Simple effective {L}agrangian for hard
  thermal loops}, {\it Phys. Rev.\/} {\bf D45}, 1827 (1992).

\bibitem{Frenkel:1992ts}
J.~Frenkel and J.~C. Taylor, \enquote{Hard thermal {QCD}, forward scattering
  and effective actions}, {\it Nucl. Phys.\/} {\bf B374}, 156 (1992).

\bibitem{Nair:1994xs}
V.~P. Nair, \enquote{Hamiltonian analysis of the effective action for hard
  thermal loops in {QCD}}, {\it Phys. Rev.\/} {\bf D50}, 4201 (1994).

\bibitem{Blaizot:1994be}
J.~P. Blaizot and E.~Iancu, \enquote{Soft collective excitations in hot gauge
  theories}, {\it Nucl. Phys.\/} {\bf B417}, 608 (1994).

\bibitem{Blaizot:1994am}
J.-P. Blaizot and E.~Iancu, \enquote{Energy momentum tensors for the
  quark-gluon plasma}, {\it Nucl. Phys.\/} {\bf B421}, 565 (1994).

\bibitem{Kelly:1994dh}
P.~F. Kelly, Q.~Liu, C.~Lucchesi and C.~Manuel, \enquote{Classical transport
  theory and hard thermal loops in the quark-gluon plasma}, {\it Phys. Rev.\/}
  {\bf D50}, 4209 (1994).

\bibitem{Bjorken:1983qr}
J.~D. Bjorken, \enquote{Highly relativistic nucleus-nucleus collisions: The
  central rapidity region}, {\it Phys. Rev.\/} {\bf D27}, 140 (1983).

\bibitem{Baym:1984np}
G.~Baym, \enquote{Thermal equilibration in ultrarelativistic heavy ion
  collisions}, {\it Phys. Lett.\/} {\bf B138}, 18 (1984).

\bibitem{Mueller:1999pi}
A.~H. Mueller, \enquote{The {B}oltzmann equation for gluons at early times
  after a heavy ion collision}, {\it Phys. Lett.\/} {\bf B475}, 220 (2000).

\bibitem{Lappi:2006fp}
T.~Lappi and L.~McLerran, \enquote{{Some features of the glasma}}, {\it Nucl.
  Phys.\/} {\bf A772}, 200 (2006).

\bibitem{Fukushima:2006ax}
K.~Fukushima, F.~Gelis and L.~McLerran, \enquote{{Initial singularity of the
  little bang}}, {\it Nucl. Phys.\/} {\bf A786}, 107 (2007).

\bibitem{Kovchegov:2000hz}
Y.~V. Kovchegov, \enquote{Classical initial conditions for ultrarelativistic
  heavy ion collisions}, {\it Nucl. Phys.\/} {\bf A692}, 557 (2001).

\bibitem{Lappi:2007ku}
T.~Lappi, \enquote{{Wilson line correlator in the MV model: relating the glasma
  to deep inelastic scattering}}, arXiv:0711.3039 [hep-ph] (2007).

\bibitem{Eskola:2005ue}
K.~J. Eskola {\it et~al.\/}, \enquote{{{RHIC}-tested predictions for low-p({T})
  and high-p({T}) hadron spectra in nearly central {Pb + Pb} collisions at the
  {LHC}}}, {\it Phys. Rev.\/} {\bf C72}, 044904 (2005).

\bibitem{Blaizot:2006tk}
J.~P. Blaizot, E.~Iancu, U.~Kraemmer and A.~Rebhan, \enquote{{Hard-thermal-loop
  entropy of supersymmetric Yang-Mills theories}}, {\it JHEP\/} {\bf 06}, 035
  (2007).

\bibitem{McLerran:1993ni}
L.~D. McLerran and R.~Venugopalan, \enquote{Computing quark and gluon
  distribution functions for very large nuclei}, {\it Phys. Rev.\/} {\bf D49},
  2233 (1994).

\bibitem{Baier:2002bt}
R.~Baier, A.~H. Mueller, D.~Schiff and D.~T. Son, \enquote{Does parton
  saturation at high density explain hadron multiplicities at {RHIC?}}, {\it
  Phys. Lett.\/} {\bf B539}, 46 (2002).

\bibitem{Krasnitz:2001ph}
A.~Krasnitz and R.~Venugopalan, \enquote{Small x physics and the initial
  conditions in heavy ion collisions}, {\it Nucl. Phys.\/} {\bf A698}, 209
  (2002).

\bibitem{Krasnitz:2003jw}
A.~Krasnitz, Y.~Nara and R.~Venugopalan, \enquote{Classical gluodynamics of
  high energy nuclear collisions: An erratum and an update}, {\it Nucl.
  Phys.\/} {\bf A727}, 427 (2003).

\end{thebibliography}

\end{document}